\documentclass[12pt,preprint]{aastex}

\begin{document}
\shorttitle{Flux Evolution shaped by the Curvature Effect}
\shortauthors{Lin et al.}
\title{Steep Decay Phase Shaped by the Curvature Effect. I. Flux Evolution}
\author{Da-Bin Lin\altaffilmark{1,2}, Hui-Jun Mu\altaffilmark{1,2,3,4}, Rui-Jing Lu\altaffilmark{1,2}, Tong Liu\altaffilmark{3,4}, Wei-Min Gu\altaffilmark{3,4}, Yun-Feng Liang\altaffilmark{5}, Xiang-Gao Wang\altaffilmark{1,2}, and En-Wei Liang\altaffilmark{1,2}
}
\altaffiltext{1}{GXU-NAOC Center for Astrophysics and Space Sciences, Department of Physics, Guangxi University, Nanning 530004, China; lindabin@gxu.edu.cn}
\altaffiltext{2}{Guangxi Key Laboratory for Relativistic Astrophysics, the Department of Physics, Guangxi University, Nanning 530004, China}
\altaffiltext{3}{Department of Astronomy, Xiamen University, Xiamen, Fujian 361005, China}
\altaffiltext{4}{SHAO-XMU Joint Center for Astrophysics, Xiamen University, Xiamen, Fujian 361005, China}
\altaffiltext{5}{Purple Mountain Observatory, Chinese Academy of Sciences, Nanjing 210008, China}
\begin{abstract}
The curvature effect may be responsible for the steep decay phase observed in gamma-ray bursts.
For testing the curvature effect with observations,
the zero time point $t_0$ adopted to
plot observer time and flux on a logarithmic scale should be appropriately selected.
In practice, however, the true $t_0$ cannot be directly constrained from the data.
Then, we move $t_0$ to a certain time in the steep decay phase, which can be easily identified.
In this situation, we derive an analytical formula to describe the flux evolution of the steep decay phase.
The analytical formula is read as
$F_\nu\propto (1+\tilde t_{\rm obs}/{\tilde t_c})^{-\alpha}$
with $\alpha(\tilde{t}_{\rm obs})=2+{\int_{0}^{\log (1+\tilde{t}_{\rm obs}/{\tilde{t}_c})} {\beta(\tau)d[\log(1+\tau/{\tilde{t}_c})]}}/{\log (1 + {\tilde t}_{\rm obs}/{{\tilde t}_c})}$,
where $F_\nu$ is the flux observed at frequency $\nu$,
$\tilde t_{\rm obs}$ is the observer time by setting zero time point $t_0$ at a certain time in the steep decay phase,
$\beta$ is the spectral index estimated around $\nu$,
and ${\tilde t}_c$ is the decay timescale of the phase with $\tilde{t}_{\rm obs}{\geqslant}0$.
We test the analytical formula with the data from numerical calculations.
It is found that the analytical formula presents a well estimation about the evolution of flux shaped by the curvature effect.
Our analytical formula can be used to confront the curvature effect with observations
and estimate the decay timescale of the steep decay phase.
\end{abstract}
\section{Introduction}\label{Sec:Introduction}
Gamma-ray bursts (GRBs) are the most powerful explosive events in the Universe.
They are always traced by the Burst Alert Telescope (BAT) in the $\gamma$-ray energy bands (\citealp{Barthelmy2005SSRv}). This phase is the so-called $\gamma$-ray prompt emission, which can last
from ten milliseconds to several minutes, and even longer
(\citealp{Kouveliotou1993,Gendre2013,Virgili2013,Stratta2013,Levan2014,Zhang2014}).
Following the $\gamma$-ray prompt emission is a long-lived afterglow emission,
which emits mainly at longer wavelengths, such
as X-ray, optical, and radio.
The observations of \emph{Swift} satellite reveal that the light curve of
X-ray afterglow emission is composed of five components (\citealp{Zhang2006,Nousek2006,OBrien2006,Zhang2007}).
The first of the these components is the initial steep decay phase,
which appears at around $10^2-10^3$ seconds after the burst trigger
(\citealp{Vaughan2006,Cusumano2006, OBrien2006}).
By extrapolating the prompt $\gamma$-ray light curve to X-ray band,
it is found the initial steep decay phase observed can connect smoothly to this extrapolated X-ray light curve.
Thus, it is believed that the initial steep decay phase may be the ``tail'' of the prompt emission
(\citealp{Barthelmy2005, OBrien2006, Liang2006}).
Beside the prompt emission phase,
the steep decay is also observed in the decay phase of flares (e.g., \citealp{Mu2016}; \citealp{Uhm2016}; \citealp{Jia2015}).
The behavior of steep decay phase is our focus in this work.

For the steep decay phase, the temporal decay index $\alpha$ of the observed flux
is typically $\sim 3-5$.
Moreover, the value of $\alpha$ is found to be correlated with the spectral index $\beta$.
This led to the development of the ``curvature effect'' model,
which plays an important role in shaping
the flux decline in the steep decay phases
(\citealp{Zhang2006,Liang2006,Wu2006,Yamazaki2006}).
When emission in a spherical relativistic jet ceases/decays abruptly,
the observed flux is controlled by high latitude's emission in the jet shell.
In this situation, the photons from higher latitude would be observed later and has a lower Doppler factor.
Then, the observed flux would progressively decrease.
For an intrinsic spectrum described as a single power-law form
(i.e., $F'\propto \nu'^{-\beta}$ with $\beta=\rm constant$ in the comoving frame of the jet shell),
the relation between $\alpha$ and $\beta$ due to the curvature effect can be read as
(see \citealp{Uhm2015} for details; \citealp{Kumar2000,Dermer2004,Dyks2005})
\begin{equation}\label{Introdution}
\alpha=2+\beta.
\end{equation}
As showed in \cite{Nousek2006}, above relation is in rough agreement
with the data on the steep decay phase of some \emph{Swift} bursts.
Adopting a time-averaged $\beta$ in the steep decay phases,
\cite{Liang2006} finds that Equation~(\ref{Introdution}) is generally valid.

For testing Equation~(\ref{Introdution}) with observations,
the zero time point ``$t_0$'' is usually discussed (\citealp{Zhang2006}; \citealp{Liang2006}; \citealp{Uhm2016}).
If one wants to find the relation as Equation~(\ref{Introdution}) based on the observational data,
the time $t_0$ for the steep decay phase should be appropriately selected.
This is because the light curves of GRB are plotted on a logarithmic scale for
both the observer time and the flux in order to find the decay slope $\alpha$.
The different reference time $t_0$ adopted to plot the light curves
would affect the obtained value of $\alpha$.
For a spherical relativistic jet moving with a constant Lorentz factor,
Equation~(\ref{Introdution}) can be found by setting $t_0$ at the observed time of jet ejection (\citealp{Uhm2015}).
In practice, however, the true $t_0$ cannot be directly constrained from the data.
The reasons are three:
(1) the radiation of jet in GRBs always begins at the radius $r_0 \gg 0$ rather than $r_0 = 0$;
(2) the detector misses the initial portion of jet radiation due to detector sensitivity;
(3) the initial portion of jet radiation may be buried under the background (\citealp{Uhm2016}).
One practical way to test the curvature effect model
may be to move $t_0$ to a certain time in the steep decay phase,
which can be easily identified in the GRB light curves.
Since the setting about $t_0$ in this situation is not the physically-motivated one,
the $\alpha-\beta$ law and even the flux evolution pattern naturally deviate from the standard law
as shown in Equation~(15) of \cite{Uhm2015}.
Then, we try to derive the analytical formula to describe the flux evolution in this situation.

The paper is organized as follows.
Since our analytical formula of flux evolution will be tested with the data from the numerical calculations,
the numerical procedures in our numerical calculations are presented in Section~\ref{Sec:Numerical calculation Procedure}.
By moving $t_0$ to a certain time in the steep decay phase,
the analytical formula of flux evolution is presented and tested
in Sections~\ref{Sec:Analytical Formula of Flux Evolution} and \ref{Sec:Testing}, respectively.
Our conclusions are summarized in Section~\ref{Sec:Conclusion}.

\section{Procedures for Simulating Jet Emission}\label{Sec:Numerical calculation Procedure}
The curvature effect is a combination of the time delay
and the Doppler shifting of the intrinsic spectrum for
high latitude emission with respect to the light of sight.
Then, the arrival time of photons and Doppler shifting of the intrinsic spectrum should be prescribed.
For an expanding spherical thin jet shell, the shell is assumed to locate at radius $r$ at time $t$,
where the value of $r$ is measured with respect to the jet base.
In addition, we discuss a spherical thin jet shell
radiating from radius $r_0$ to $r_e$.
Thus, the arrival time for photons from an emitter in the jet shell to observer is
\begin{equation}\label{Eq:t_obs}
{t_{{\rm{obs}}}} =\left\{ \int_{{r_0}}^{{r}} {[1 - \beta_{\rm jet}(l)]} \frac{{dl}}{c\beta_{\rm jet}(l) }+ \frac{r(1- \cos \theta)}{c}\right\}(1+z),
\end{equation}
where the emitter locates at ($r,\,\theta$),
$c\beta_{\rm jet}(l)=cdr/dt$ is the velocity of jet shell at radius $r=l$, $c$ is the light velocity,
$\theta$ is the polar angle of the emitter with respect to the line of sight in spherical coordinates
(the origin of coordinate is at the jet base),
and $z$ is the redshift of the explosion producing the jet shell.
In Equation~(\ref{Eq:t_obs}), $t_{\rm obs}=0$ is set at the observed time for the first photon,
which is from the emitter located at $r=r_0$ and $\theta=0$.
In addition, we define
\begin{equation}\label{Eq:t_obs_r}
t_{{\rm obs},r}\equiv (1 + z)\int_{{r_0}}^r {[1-\beta_{\rm jet}(l)]}\frac{dl}{c\beta_{\rm jet}(l)},
\end{equation}
which is the observed time for first photon from jet shell located at $r$.

The radiation of an electron is always discussed with relativistic electrons ($\gamma'_e\;\gg 1$) and strong magnetic field.
With these two ingredients,
photons are produced by synchrotron process, and can be scattered to higher energy by inverse Compton process.
In our work, the shape of radiation spectrum is important rather than the detailed radiation processes.
To simplify the problem, the radiation spectrum of an electron with $\gamma'_e$ is assumed as (e.g., \citealp{Uhm2015})
\begin{equation}\label{eq:spec_power_single_ensemble}
P'_{\nu'}(\nu') =P'_0\, H'(\nu'/\nu'_0),
\end{equation}
where $P_0^{\prime}$ describes the spectral power in the jet shell comoving frame
and $\nu'_0$ is the characteristic radiation frequency.
The values of $P'_0$ and $\nu'_0$ may be related to $\gamma'_e$ and thus may evolve with time.
It should be noted that the above description about the radiation spectrum follows that of \cite{Uhm2015}.
For the analytical formula of $H'(x)$, we study following four cases:
\begin{equation}\label{Eq:H function}
\begin{array}{*{20}{c}}
{({\rm{I}}):}&{H'(x) = {x^{ - {{\hat \beta '}_0} - k\log (x)}},}\\
{({\rm{II}}):}&{H'(x){\rm{ = }}{x^{\alpha '+1}}{{\left[ {1 + g\,{x^{(\alpha '- \beta ')w}}} \right]}^{ -1/w}},}\\
{({\rm{III}}):}&{H'(x) = \left\{ {\begin{array}{*{20}{c}}
{{x^{\alpha ' + 1}}\exp \left( { - x} \right),}&{x \le \left( {\alpha ' - \beta '} \right),}\\
{{{\left( {\alpha ' - \beta '} \right)}^{\alpha ' - \beta '}}\exp \left( {\beta ' - \alpha '} \right){x^{\beta ' + 1}},}&{x \ge \left( {\alpha ' - \beta '} \right),}
\end{array}} \right.}\\
{({\rm{IV}}):}&{H'(x) = x^3\left[\exp (x) - 1\right]^{-1},}
\end{array}
\end{equation}
where $\hat \beta'_0$, $k$, $\alpha'$, $\beta'$, $w=2$ and $g=6.8$ are constants.
Case (III) is used to discuss the situations with a ``Band-function'' intrinsic spectrum (\citealp{Band1993}),
which is a two joint functions. For this case, the spectral slope in the space of ``$\log \nu' -\log P'_{\nu'}$'',
i.e., ${\rm d}\log P'_{\nu'}/{\rm d}\log \nu'$, may be peculiar  at frequency $\nu'\sim ({\alpha ' - \beta '})\nu'_0$.
This behavior can be found in Section~\ref{Sec:Testing}.
Owing to this behavior,
we introduce a smooth joint broken power-law function, i.e., Case (II),
to mimic a ``Band-function'' spectrum.
The spectral evolution in this case is smoother than that in Case (III).
In general, most of radiation spectra, such as Cases (II) and (III), can be described as
$H'(\nu'/\nu'_0)\propto (\nu'/\nu'_1)^{-\hat \beta'(\nu'_1, \nu')}$ with a $\nu'_1$- and $\nu'$-dependent $\hat \beta '$.
Moreover, the spectral slope of most radiation spectra in the ``$\log \nu' -\log P'_{\nu'}$'' space varies slowly.
Then, we can use $\hat \beta'(\nu'_1, \nu')=\hat \beta'(\nu'_1, \nu'_2)+k\log(\nu'/\nu'_2)$
to approximately describe $\hat{\beta}'(\nu'_1, \nu')$ for different $\nu'$ and $\log(\nu'/\nu'_2)\sim 0$.
In this situation, we have $H'(\nu'/\nu'_0)\propto (\nu'/\nu'_1)^{-\hat \beta'(\nu'_1, \nu')}
\approx(\nu'/\nu'_1)^{-\hat \beta'_0-k\log(\nu'/\nu'_1)}$, i.e., Case (I),
where $\hat \beta'_0=\hat \beta'(\nu'_1, \nu'_2)+k\log(\nu'_1/\nu'_2)$.
In this paper, Case (I) is used to illustrate the features of steep decay phase
shaped by the shell curvature effect.
The observed $\gamma$-ray prompt emission of GRBs may be from the photosphere,
of which the radiation spectra may be similar to the spectrum of blackbody radiation, i.e., Case (IV).
Owing to the complication in modeling the jet dynamic for the photosphere emission,
however, we only discuss Case (IV) for an extreme fast cooling thin shell in this work (see Section~\ref{Sec:Testing_EFCSs}).
It should be noted that the photospheric surface is not spherical (\citealp{Pe'er_A-2008}; \citealp{Beloborodov_AM-2011}; see \citealp{Deng2014} and references therein).

In our numerical calculations, the radiation of jet shell at time $t$
are modelled with a number of emitters randomly distributed in the jet shell.
The number of relativistic electrons $n'_e$ is the same for different emitters.
Thus, the total radiation power from an emitter in the comoving frame is $n'_eP'_0\, H(\nu'/\nu'_0)$,
which is the same for different emitters.
For a relativistic moving jet with a Lorentz factor $\Gamma$,
the comoving emission frequency $\nu'$ is boosted to $\nu=D\nu'$ in the observer's frame.
Here, $D$ is the Doppler factor described as
\begin{eqnarray}\label{}
D ={\left[ {{\Gamma }(1 - {\beta _{{\rm{jet}}}}\cos \theta )} \right]^{ - 1}}.
\end{eqnarray}
During the shell's expansion for $\delta t\; (\sim 0)$,
the observed spectral energy $\delta U$ from an emitter
into a solid angle $\delta \Omega$ in the direction of the observer is given as (\citealp{Uhm2015})
\begin{equation}\label{Eq:Numerical Spectral_Energy}
\delta {U_\nu} (t_{\rm obs}) = \left({D^2}\delta \Omega\right)\left( \frac{{\delta t}}{\Gamma }\right)\frac{1}{{4\pi }}{n'_e}{P'_0}H'\left( {\frac{{\nu(1 + z)}}{{D{\nu'_0}}}} \right)
\end{equation}
where the emission of electrons is assumed isotropically in the jet shell comoving frame (c.f. \citealp{Geng_JJ-2017-Huang_YF}).

The procedures for obtaining the observed flux is shown as follows.
Firstly, an expanding jet is modelled with a series of jet shells
at radius $r_0,\;r_1=r_0+\beta_{\rm jet}(r_0)c{\delta t},\;r_2=r_1+\beta_{\rm jet}(r_1)c{\delta t},\;\cdot\cdot\cdot,r_n=r_{n-1}+\beta_{\rm jet}(r_{n-1})c{\delta t},\;\cdot\cdot\cdot$
appearing at the time $t=0{\rm s},\;{\delta t},\;{2\delta t},\;\cdot\cdot\cdot, n{\delta t},\;\cdot\cdot\cdot $
with velocity $c\beta_{\rm jet}(r_0),\;c\beta_{\rm jet}(r_1),\;c\beta_{\rm jet}(r_2),\;\cdot\cdot\cdot,c\beta_{\rm jet}(r_n),\;\cdot\cdot\cdot$, respectively.
During the shell's expansion for $\delta t$,
the shell move from $r_{n-1}$ to $r_n$ with the same radiation behavior for emitters.
Secondly, we produce $N$ emitters centred at ($r_n$, $\theta$, $\varphi$) in spherical coordinates,
where the value of $\cos\theta$ and $\varphi$ are randomly picked up from linear space of $[\cos\theta_{\rm jet},1]$ and $[0,2\pi]$, respectively. Here, $\theta_{\rm jet}$ is the half-opening angle of jet.
The observed spectral energy from an emitter
during the shell's expansion from $r_{n-1}$ to $r_n$ is calculated with Equation~(\ref{Eq:Numerical Spectral_Energy}).
By discretizing the observer time $t_{\rm obs}$ into a series of time intervals,
i.e., $[0, {\delta t_{\rm obs}}],\,[{\delta t_{\rm obs}},2{\delta t_{\rm obs}}]\, \cdot\cdot\cdot\,,[(k-1){\delta t_{\rm obs}}, k{\delta t_{\rm obs}}],\cdot\cdot\cdot$,
we can find the total observed spectral energy
\begin{equation}
\left. U_\nu \right|_{\left[(k - 1)\delta {t_{{\rm{obs}}}},k\delta {t_{{\rm{obs}}}} \right)}=
\sum\limits_{(k-1)\delta t_{\rm obs}\leqslant t_{\rm obs}<k\delta t_{\rm obs}}{\delta {U_\nu} (t_{\rm obs})}
\end{equation}
in the time interval $[(k-1){\delta t_{\rm obs}}, k{\delta t_{\rm obs}}]$ based on Equations~(\ref{Eq:t_obs}) and (\ref{Eq:Numerical Spectral_Energy}). Here, $t_{{\rm obs},r_n}=\sum\limits_{i=0}^{n-1}{[1-\beta_{\rm jet}(r_i)]\delta t}$ and $t_{{\rm obs},r_0}=0$ are used. Then, the observed flux at the time $(k/2-1){\delta t_{\rm obs}}$ is
\begin{equation}\label{Eq:Numerical calculation-Flux}
F_\nu=\frac{{{{\left. U_\nu \right|}_{\left[(k - 1)\delta {t_{{\rm{obs}}}},k\delta {t_{{\rm{obs}}}} \right)}}}}{{D_{\rm{L}}^2\delta {t_{{\rm{obs}}}}\delta \Omega}},
\end{equation}
where $D_{\rm L}$ is the luminosity distance of the jet shell with respect to the observer.
In our numerical calculations,
the jet shell is assumed to begin radiation at radius $r_0=10^{14}\rm cm$
with a Lorentz factor $\Gamma(r_0)=\Gamma_0=300$.
The value of $\nu'_0=1\rm keV(1+z)$, $N\gg 1$, $\theta_{\rm jet}\gg 1/\Gamma_0$,
$\delta t << t_{c,r_0}$, and $\delta t_{\rm obs}=0.005t_{c,r_0}$
are adopted and remained as constants in a numerical calculation, where $t_{c,r}=r(1+z)/\Gamma^2c$.
The total duration of our producing light curves are set as $50t_{c,r_0}$.
Then, the obtained data would be significantly large.
To reduce the file size of our figures, we only plot the data in the time interval with $k$
satisfying $(k-1){\delta t_{\rm obs}}<1.1^m\times 0.01t_{c,r_0}+{t_0}<k{\delta t_{\rm obs}}$,
where $m\;(\geqslant 0)$ is an any integer and ${t_0}$ is the observer time set for $\tilde{t}_{\rm obs}=0$ (see Section~\ref{Sec:Analytical Formula of Flux Evolution}).
The light curves in these figures are consistent with those plotted based on all of data from our numerical calculations.

\section{Analytical Formula of Flux Evolution}\label{Sec:Analytical Formula of Flux Evolution}
In this section,
the analytical formula of flux evolution is obtained by analyzing the radiation from an extreme fast cooling thin shell (EFCS).
For this situation, we assume the radiation behavior of jet shell unchanged during the shell's expansion time ${\delta t}\sim 0$.
Then, we have $r= r_0 + c\beta_{\rm jet}c{\delta t}\sim r_0$, and Equation~(\ref{Eq:t_obs}) can be reduced to
\begin{equation}\label{Eq:t_obs_thin}
t_{\rm obs}=(r/c)(1-\cos\theta)(1+z),
\end{equation}
which describes the delay time of photons from ($r,\,\theta$) with respect to those from ($r,\,\theta=0$).
It reveals that $t_{\rm obs}= 0$ is the beginning of the phase shaped by the curvature effect
for flux from an EFCS.
That is to say, $t_{\rm obs}$ is the observer time by setting zero time point $t_0$ at the beginning of the steep decay phase.
Then, our obtained analytical formula for the flux evolution in the steep decay phase,
i.e., Equation~(\ref{Eq:Fnu_Evolution}),
describes the flux evolution by setting $t_0$ (i.e., $t_{\rm obs}= 0$)
at the beginning of the phase shaped by the shell curvature effect.
With $\Gamma\gg 1$, $D$ can be reduced to
\begin{equation}
D \approx {\left\{ {\Gamma  - \Gamma \left( {1 - \frac{1}{2\Gamma ^2}} \right)\left [1 - (1 - \cos \theta )\right ]} \right\}^{ - 1}} \approx {\left[ {\frac{1}{{2\Gamma }} + \Gamma (1 - \cos \theta )} \right]^{ - 1}},
\end{equation}
or
\begin{equation}\label{Eq:D}
D\approx \frac{2\Gamma}{1+t_{\rm obs}/t_{c,r}},
\end{equation}
where $t_{c,r}$ is the characteristic timescale of shell curvature effect at the radius $r$,
\begin{equation}\label{Eq:Gamma}
t_{c,r}=\frac{r(1 + z)}{2\Gamma ^2c}.
\end{equation}
The difference between $D$ and ${2\Gamma}/{(1+t_{\rm obs}/t_{c,r})}$ can be neglected for significantly large value of $\Gamma$. Then, we would like to use $D={2\Gamma}/{(1+t_{\rm obs}/t_{c,r})}$ in our analysis.

For the observed time interval $\delta t_{\rm obs}$,
the observed total number of emitter is $N|\delta (\cos\theta)|/(1-\cos\theta_{\rm jet})$
with $|\delta (\cos\theta)|=c\delta t_{\rm obs}/r(1+z)$ derived based on Equation~(\ref{Eq:t_obs_thin}).
Then, the observed flux at the time $t_{\rm obs}$ is
\begin{equation}
F_\nu=\frac{\delta {U_\nu} (t_{\rm obs})N|\delta (\cos\theta)|/(1-\cos\theta_{\rm jet})}{D_{\rm{L}}^2\delta t_{\rm obs}\delta \Omega},
\end{equation}
or,
\begin{equation}\label{Eq:Observed flux}
F_\nu=AD^2H'(\nu(1+z)/D\nu'_0),
\end{equation}
where $A={n'_e}{P'_0}Nc\delta t/[4\pi D_{\rm L}^2\Gamma r(1 - \cos {\theta _{{\rm{jet}}}})(1 + z)]$ is a constant for an EFCS.
The method used to derive Equation~(\ref{Eq:Observed flux}) is from \cite{Uhm2015}.
As shown in \cite{Uhm2015}, Equation~(\ref{Eq:Observed flux}) can also be derived with another method.
The reader can read the above paper for the details.
For a constant observed frequency $\nu=D\nu'/(1+z)$,
the observed flux $F_{\nu}$ in Case (I) can be described as
\begin{equation}\label{Eq:Spectral_Evolution}
{F_\nu } = F_{\nu _0}{\left( {\frac{\nu }{{{\nu _0}}}} \right)^{ - \beta }}
= F_{\nu_0}{{{\left( {\frac{\nu }{{{\nu _0}}}} \right)}^{ - \hat\beta'_0 - k\log (\nu /{\nu _0}) - 2k\log (1 + t_{\rm obs}/t_{c,r})}}}
\end{equation}
with
\begin{equation}\label{Eq:F_Evolution}
{F_{\nu_0}}={F_{\nu_0,0}}{\left(1+ \frac{t_{\rm obs}}{t_{c,r}}\right)^{ - 2 - \hat \beta'_0-k\log(1+t_{\rm obs}/t_{c,r})}},
\end{equation}
where $\nu_0=D_0\nu'_0/(1+z)$, $D_0=2\Gamma$ is the Doppler factor of the emitter observed at $t_{\rm obs}=0$,
and $F_{{\nu_0},0}$ is the observed flux at time $t_{\rm obs}=0$ and frequency $\nu_0$.
With Equations~(\ref{Eq:Spectral_Evolution}) and (\ref{Eq:F_Evolution}), the evolution of flux from an EFCS can be described as
\begin{equation}\label{Eq:Fnu_Evolution}
{F_\nu } = {F_{\nu ,0}}{\left( {1 + \frac{t_{\rm obs}}{{{t_c}}}} \right)^{ -\alpha(t_{\rm obs})}},
\end{equation}
where $F_{\nu ,0}$ is the flux observed at $t_{\rm obs}=0$,
and $t_c$ is the decay timescale for the phase shaped by shell curvature effect.
The main ingredients of Equation~(\ref{Eq:Fnu_Evolution}) are the value of $t_c$ and temporal decay index $\alpha (t_{\rm obs})$.
For the value of $\alpha$, it is always associated with the spectral index $\beta$.
Based on the discussion in Appendix~\ref{App:alpha_beta_relation}, we have
\begin{equation}\label{Eq:alpha_beta_relation}
\alpha(t_{\rm obs})=2+\frac{1}{\log (1 + t_{\rm obs}/t_c)}\int_{0}^{{\log (1 + t_{\rm obs}/t_c)}} {\beta(\tau)d[\log (1 + \tau/t_c)]}.
\end{equation}
It should be noted that
for flux from an EFCS located at $r_0$, $t_{\rm obs}= 0$ is the beginning of the phase shaped by the curvature effect.
Then, Equation~(\ref{Eq:Fnu_Evolution})
with $\alpha$ evolving as Equation~(\ref{Eq:alpha_beta_relation})
describes the flux evolution by setting $t_0$ at the beginning of the steep decay phase.

In practice, one may set $t_0$ at a certain time in the steep decay phase rather than the beginning of the steep decay phase.
We take ${t_p^0} (>0)$ as the time difference of $t_0$ with respect to the beginning of the steep decay phase.
By defining $\tilde{t}_{\rm obs}=t_{\rm obs}-{t_p^0}$,
Equation~(\ref{Eq:D}) is reduced to
\begin{equation}\label{Eq:D_t0}
D(\tilde{t}_{\rm obs})=\frac{2\Gamma}{1+{t_p^0}/t_{c,r}}\frac{1}{1+\tilde{t}_{\rm obs}/(t_{c,r}+{t_p^0})}
=\frac{D_0}{1+\tilde{t}_{\rm obs}/\tilde{t}_{c,r}},
\end{equation}
where $D_0={2\Gamma}/({1+{t_p^0}/t_{c,r}})$ is the Doppler factor of emitter observed at $\tilde{t}_{\rm obs}=0$
and $\tilde{t}_{c,r}=t_{c,r}+{t_p^0}$ is adopted.
With the same process to derive Equations~(\ref{Eq:Fnu_Evolution}) and (\ref{Eq:alpha_beta_relation}),
one can have
\begin{equation}\label{Eq:Fnu_Evolution_t0}
F_\nu(\tilde{t}_{\rm obs}) = \tilde{F}_{\nu ,0}{\left(1 + \frac{\tilde{t}_{\rm obs}}{\tilde{t}_c}\right)^{-\alpha(\tilde{t}_{\rm obs})}}
\end{equation}
with
\begin{equation}\label{Eq:alpha_beta_relation_t0}
\alpha(\tilde t_{\rm obs})=2+\frac{\int_{0}^{\log (1 + {\tilde t}_{\rm obs}/{{\tilde t}_c})} {\beta (\tau)d[\log (1 + \tau/{{\tilde t}_c})]}}{\log (1 + {\tilde t}_{\rm obs}/{{\tilde t}_c})},
\end{equation}
where $\tilde{F}_{\nu ,0}$ is the observed flux at $\tilde{t}_{\rm obs}=0$
and ${\tilde t}_c=t_c+{t_p^0}$ (see Appendix~\ref{Sec:tc_discussion} for the discussion about the value of $\tilde{t}_c$) is the decay timescale for the phase with $\tilde{t}_{\rm obs}\geqslant 0$.

\emph{Equation~(\ref{Eq:Fnu_Evolution_t0})
is our obtained analytical formula about the flux evolution
in the steep decay phase,
where the zero time point $t_0$ ($\tilde t_{\rm obs}=0$) is set at a certain time in the steep decay phase.}
We will test Equation~(\ref{Eq:Fnu_Evolution_t0}) in different situations
with $\tilde{F}_{\nu ,0}$ estimated based on the numerical calculations' data at $\tilde t_{\rm obs}=0$.

\section{Testing the Analytical Formula of Flux Evolution }\label{Sec:Testing}

\subsection{Testing with EFCSs}\label{Sec:Testing_EFCSs}
First, we test Equation~(\ref{Eq:Fnu_Evolution_t0}) in the situation with an EFCS and Case (I).
By setting $t_0=0$, we have $\tilde{t}_{\rm obs}=t_{\rm obs}$ and $\tilde{t}_c=t_{c,r}$.
Equation~(\ref{Eq:Spectral_Evolution}) reveals the following relation:
\begin{equation}\label{Eq:Beta_Evolution}
\beta(\tilde t_{\rm obs})={\beta_0}+ 2k\log (1 + \tilde t_{\rm obs}/{\tilde t_{c}}),
\end{equation}
where $\beta=-[\log(F_{\nu_1})-\log(F_\nu)]/[\log(\nu_1)-\log(\nu)]$ with $\nu_1\sim \nu$ is used and
$\beta_0$ is the value of $\beta$ at $\tilde t_{\rm obs}=0$.
Since the value of $\beta$ is estimated around $\nu$, the relation of
\begin{equation}\label{Eq:alpha}
\alpha(\tilde t_{\rm obs})=2+{\beta_0}+ k\log (1 + \tilde t_{\rm obs}/\tilde t_{c})
\end{equation}
can be found based on Equations~(\ref{Eq:Spectral_Evolution}) and (\ref{Eq:F_Evolution}),
where
\begin{equation}
{(\nu /{\nu _0})^{2k\log (1 + {\tilde t_{{\rm{obs}}}}/\tilde t_{c})}} = {(1 + {\tilde t_{{\rm{obs}}}}/\tilde t_{c})^{2k\log (\nu /{\nu _0})}}
\end{equation}
and $\hat{\beta}'_0+2k\log (\nu /{\nu _0})=\beta_0$ are used.
Above analysis shows that the flux evolution in the situation with an EFCS and Case (I) can be described as
\begin{equation}\label{Eq:Testing}
F_\nu(\tilde{t}_{\rm obs}) = \tilde{F}_{\nu ,0}{\left(1 + \frac{\tilde{t}_{\rm obs}}{\tilde{t}_c}\right)^{-2-{\beta_0}- k\log (1 + \tilde t_{\rm obs}/\tilde t_{c})}}.
\end{equation}
It is important to point out that Equation~(\ref{Eq:Testing}) can also be derived based on Equation~(\ref{Eq:Fnu_Evolution_t0}).
According to Equations~(\ref{Eq:alpha_beta_relation_t0}) and (\ref{Eq:Beta_Evolution}),
we have
\begin{equation}
\alpha=2+\frac{\int_{0}^{\log (1 + \tilde t_{\rm obs}/{\tilde t_{c}})} {[{\beta_0}+ 2k\log (1 + \tau/{\tilde t_{c}})]d[\log (1 + \tau/\tilde t_c)]}}{\log (1 + \tilde t_{\rm obs}/\tilde t_c)}=2+{\beta_0}+ k\log (1 + \tilde t_{\rm obs}/{\tilde t_{c}}).
\end{equation}
Substituting above relation into Equation~(\ref{Eq:Fnu_Evolution_t0}), one can have
\begin{equation}\label{Eq:alpha_beta_derivation}
F_\nu(\tilde{t}_{\rm obs}) = \tilde{F}_{\nu ,0}{\left(1 + \frac{\tilde{t}_{\rm obs}}{\tilde{t}_c}\right)^{-2-{\beta_0}- k\log (1 + \tilde t_{\rm obs}/\tilde t_{c})}},
\end{equation}
which is the same as Equation~(\ref{Eq:Testing}).
This reveals that Equations~(\ref{Eq:Fnu_Evolution_t0}) describes the flux evolution in the steep decay phase
shaped by an EFCS with Case (I).
It should be noted that Equations~(\ref{Eq:Beta_Evolution})-(\ref{Eq:alpha_beta_derivation}) is also applicable for ${t_p^0}\neq 0$ (i.e., $t_0>0$).

We test Equations~(\ref{Eq:Fnu_Evolution_t0}) and (\ref{Eq:Beta_Evolution}) in Figure~1,
which shows the evolution of flux (upper panels)
and spectral indexes (lower panels) for an EFCS with Case (I).
Here, $\hat \beta'_0=0$ and $k=0.5$ are adopted in the numerical calculations.
The red ``$\times$'' and violet ``$\circ$'' represent the data
for observed photon energy $h\nu=300{\rm keV}$ and $900{\rm keV}$, respectively.
In addition, $t_0=0\rm s$ and $10t_{c,r_0}$ are adopted in the left and right panels, respectively.
For comparison, we plot Equations~(21) and (23) with red (violet) solid lines
in the upper and lower panels for $h\nu=300{\rm keV}$ ($900{\rm keV}$), respectively.
Here, the values of $\tilde{F}_{\nu ,0}$ and $\beta _0$ are estimated based on the data at $\tilde t_{\rm obs}=0$,
and $\tilde t_c=t_{c, r_0}$ ($11t_{c,r_0}$) is adopted in the left (right) panels.
From this figure, one can find that Equation~(\ref{Eq:Beta_Evolution})
well describes the spectral evolution for the radiation from an EFCS with Case (I).
Moreover, Equation~(\ref{Eq:Fnu_Evolution_t0})
can describe the flux evolution in the steep decay phase.
It should be noted that the value of $F_\nu$ is almost constant for $\tilde t_{\rm obs}/\tilde t_c<<1$.
This behavior can be found in our figures and could not be read as $\alpha=0$.

In reality, the intrinsic radiation spectrum may be similar to that of Case (II) or (III).
Then, we study the radiation behavior for an EFCS with Case (II) or (III).
The evolution of flux and $\beta$ are showed in Figure~2, where $\alpha'=-1$ and {\bf $\beta'=-2.3$} are adopted.
In this figure, the black ``+'', red ``$\times$'', and violet ``$\circ$''
represent the data for $h\nu=100{\rm keV}$, $300{\rm keV}$, and $900{\rm keV}$, respectively.
In this figure and afterwards, the light curves for $h\nu=300{\rm keV}$ ($900{\rm keV}$)
are shifted by dividing $1.5$ ($3$) in the plot for clarity.
The upper part in this figure is the data with ${t_0}=0$, the lower part is the data with ${t_0}=t_{c,r_0}$,
the left panels are the data from the EFCS with Case (II), and the right panels are the data from the EFCS with Case (III).
For comparison, we show Equation~(\ref{Eq:Fnu_Evolution_t0}) with solid lines in each panels,
where the black, red, and violet solid lines are for the observed photon energy $h\nu=100{\rm keV}$, $300{\rm keV}$, and $900{\rm keV}$, respectively.
In addition, $\tilde t_c=t_{c,r_0}$ and $2t_{c,r_0}$ are adopted for upper and lower part, respectively.
It can be found that the solid lines are well consistent with the numerical calculations' data.
We also study the radiation of an EFCS with Case (IV).
The results are showed in Figure~3, of which symbols and lines have the same meaning as those in Figure~2.
The flux is plotted with ${t_0}=0\rm s$ and $t_{c,r_0}$ in the left and right panels, respectively.
Equation~(\ref{Eq:Fnu_Evolution_t0}) is showed with solid lines,
where $\tilde{t}_c={t}_{c,r_0}$ and $2{t}_{c,r_0}$ are adopted for left and right panels, respectively.
It can be found that the solid lines are well consistent with the numerical calculations' data.

Then, we can conclude that Equation~(\ref{Eq:Fnu_Evolution_t0})
can present a well estimation about the evolution of flux shaped by the shell curvature effect
for the radiation of an EFCS.

\subsection{Real Situation for a Thin Shell}\label{Sec:reality}
In this subsection, Equation~(\ref{Eq:Fnu_Evolution_t0}) is tested
with a thin shell radiating from radius $r_0$ to $r_e=2r_0$.
Since the Lorentz factor of jet shell may be related to $r$ (such as \citealp{Uhm2015}, \citealp{Uhm2016}),
we assume
\begin{equation}\label{Eq:Gamma_Evolution}
\Gamma=\Gamma_0\left(\frac{r}{r_0}\right)^s,
\end{equation}
where $s>0\;(<0)$  represents an accelerating (decelerating) jet.
In addition, it is assumed that $n'_e$ increases with time $t'$ in the comoving frame of jet shell, i.e., $n'_e=n'_{e,0}t'$,
and $t'=0$ is set at the radius $r_0$.
Since the light curves are normalized by the peak flux in our focus phase,
the exact value of constant $n'_{e,0}$ does not matter in our work.
Based on Equations~(\ref{Eq:t_obs_r}) and (\ref{Eq:Gamma_Evolution}),
the observed time $t_{\rm obs, stop}=t_{{\rm obs},r_e}$ of the jet shell stopping radiation
are $t_{\rm obs, stop}=2.33t_{c,r_0}$, $t_{c,r_0}$, and $0.5t_{c,r_0}$
for $s=-1$, $0$, and $1$, respectively.

In Figure~4, we show the flux evolution for Case (I) with $k=0$ and $\hat{\beta}'_0=1.3$,
where $s=-1$, $0$, and $1$ are adopted for left, middle, and right panels, respectively.
The gray ``$\circ$'', blue ``+'', and red ``$\times$'' represent the data by
setting ${t_0}=0$, ${t_0}=t_p$, and ${t_0}=5t_{c,r_0}$, respectively.
Here, $t_p$ is the peak time of observed flux
and $t_p=2.32t_{c,r_0}$, $1.01t_{c,r_0}$, and $0.50t_{c,r_0}$ are found for $s=-1$, $0$, and $1$, respectively.
By comparing $t_p$ with $t_{\rm obs, stop}$, $t_p$ in our light curves is the observed time of the jet shell stopping radiation.
Then, the phase with $t_{\rm obs}\geqslant t_p$ is dominated by the shell curvature effect.
For the phase with $t_{\rm obs}\geqslant t_p$,
the spectral index $\beta(\tilde{t}_{\rm obs})=1.3$ is found.
Then, we fit the flux plotted as blue ``$+$'' with Equation~(\ref{Eq:Fnu_Evolution_t0}) and $\alpha=3.3$,
which is showed with blue solid lines in this figure.
The value of $\tilde t_{c}=6.41t_{c,r_0}$, $1.96t_{c,r_0}$, and $0.69t_{c,r_0}$ are reported from our fittings
for $s=-1$, $0$, and $1$, respectively.
By comparing the solid lines with the data, it can be found
that Equation~(\ref{Eq:Fnu_Evolution_t0}) can present a better estimation about the flux evolution.
We also fit the flux plotted as red ``$\times$'' with Equation~(\ref{Eq:Fnu_Evolution_t0}) and $\alpha=3.3$,
which is showed with red lines in this figure.
The value of $\tilde t_{c}=9.26t_{c,r_0}\approx (5t_{c,r_0}-t_p)+6.41t_{c,r_0}$,
$6.13t_{c,r_0}\approx (5t_{c,r_0}-t_p)+1.96t_{c,r_0}$,
and $5.39t_{c,r_0}\approx(5t_{c,r_0}-t_p)+0.69t_{c,r_0}$
are reported from our fittings for $s=-1$, $0$, and $1$, respectively.
According to Equation~(\ref{Eq:Fnu_Evolution_t0}),
the decay timescale $\tilde t_{c}$ in this situation (i.e., $t_0=5t_{c,r_0}$)
would be larger than that found in the situation with $t_0=t_p$ by $t_p^0=5t_{c,r_0}-t_p$.
The reported value of $\tilde t_{c}$ in the situation with $t_0=5t_{c,r_0}$
is consistent with Equation~(\ref{Eq:Fnu_Evolution_t0}).
Then, Equation~(\ref{Eq:Fnu_Evolution_t0}) can be used to describe the flux evolution
in the steep decay phase for a radiating thin shell with Case (I).

Figures~5 and 6 show the flux and spectral evolution for Cases (II) and (III) with $\alpha'=-1$ and $\beta'=-2.3$, respectively.
The upper, middle, and lower parts show the light curves with ${t_0}=0$,
$t_p$, and $t_p+2t_{c,r_0}$ respectively, where the value of $t_p=2.35t_{c,r_0}$, $1.01t_{c,r_0}$, and $0.50t_{c,r_0}$
are found in our numerical calculations with $s=-1$, $s=0$, and $s=1$, respectively.
The meaning of symbols and solid lines in Figures~5 and 6 are the same as those in Figure~2.
Here, the value of $\tilde t_{c}=6.41t_{c,r_0}$ ($8.41t_{c,r_0}$), $1.96t_{c,r_0}$ ($3.96t_{c,r_0}$), and $0.69t_{c,r_0}$ ($2.69t_{c,r_0}$), which are found in the numerical calculations with Case (I),
are used for $s=-1$, $0$, and $1$ with ${t_0}=t_p$ (${t_0}=t_p+2t_{c,r_0}$), respectively.
From Figures~5 and 6, one can find that Equation~(\ref{Eq:Fnu_Evolution_t0})
can present a better estimation about the flux evolution in the steep decay phase.
In Figures~7 and 8, we study the applicable of
Equation~(\ref{Eq:Fnu_Evolution_t0}) for Case (II) and (III) with $t_0=t_p$ and different ($\alpha', \beta'$),
i.e., ($\alpha'=-0.7,\beta'=-2$), ($-0.7,-2.6$), ($-1,-2$), and ($-1,-2.6$).
The meaning of symbols and lines are the same as those in Figure~2,
and the value of $\tilde t_{c}=6.41t_{c,r_0}$, $1.96t_{c,r_0}$, and $0.69t_{c,r_0}$
are used to plot solid lines following Equation~(\ref{Eq:Fnu_Evolution_t0})
for $s=-1$, $0$, and $1$, respectively.
It can be found that the solid lines present better estimation about the flux evolution in the steep decay phase.

Then, we can conclude that Equation~(\ref{Eq:Fnu_Evolution_t0})
is applicable to describe the flux evolution in the phase shaped by the shell curvature effect.

\section{Conclusions}\label{Sec:Conclusion}
For the radiation from a relativistic expanding spherical shell,
the curvature effect may play an important role
in shaping the flux evolution in the steep decay phase.
In this work, we study the steep decay phase shaped
by the shell curvature effect in details.
We move the zero time point $t_0$ to a certain time in the steep decay phase
and derive an analytical formula to describe the flux evolution in the steep decay phase.
Our obtained analytical formula is read as
$F_\nu\propto (1+\tilde t_{\rm obs}/{\tilde t_c})^{-\alpha}$
with $\alpha(\tilde t_{\rm obs})=2+{\int_{0}^{\log (1 + {\tilde t}_{\rm obs}/{{\tilde t}_c})} {\beta (\tau)d[\log (1 + \tau/{{\tilde t}_c})]}}/{\log (1 + {\tilde t}_{\rm obs}/{{\tilde t}_c})}$,
where $F_\nu$ is the observed flux at a constant observed frequency $\nu$,
$\tilde t_{\rm obs}$ is the observer time by setting $t_0$ at a certain time in the steep decay phase,
$\beta$ is the spectral index estimated around $\nu$,
and ${\tilde t}_c$ is the decay timescale of the phase with $\tilde t_{\rm obs}\geqslant 0$.
We test our analytical formula with numerical calculations.
It is found that our analytical formula can present a well estimation about the evolution of flux shaped by the curvature effect.
Our analytical formula can be used to test the curvature effect with observations
and estimate the decay timescale ${\tilde t}_c$ of the steep decay phase.

\acknowledgments
We thank the anonymous referee of this work for beneficial suggestions that improved the paper.
We also thank Bing Zhang and Dai Zi-Gao for helpful discussions.
This work is supported by the National Basic Research Program of China (973 Program, grant No. 2014CB845800),
the National Natural Science Foundation of China (Grant Nos. 11403005, 11533003, 11673006, 11573023, 11473022, 11363002, U1331202),
the Guangxi Science Foundation (Grant Nos. 2016GXNSFDA380027, 2014GXNSFBA118004, 2016GXNSFFA380006, 2014GXNSFBA118009),
and the Innovation Team and Outstanding Scholar Program in Guangxi Colleges.

\appendix
\section{Relation of $\alpha$ and $\beta$}\label{App:alpha_beta_relation}
The value of $\alpha$ is always associated with the spectral index $\beta$.
This can be found in Equation~(\ref{Eq:F_Evolution}) with $k=0$.
For this situation, we can find $\beta=\hat{\beta'_0}$ and thus $\alpha=2+\beta$.
Then, Equation~(\ref{Eq:Fnu_Evolution}) with $t_{\rm obs}\gtrsim t_c$ can be reduced to $F_\nu \propto t_{\rm obs}^{-2-\beta}$,
which has been extensively discussed in previous works (e.g., \citealp{Fenimore1996,Kumar2000,Dermer2004,Dyks2005}).
However, what would be the relation between $\alpha$ and $\beta$ if $\beta$ varies with the observer time $t_{\rm obs}$.
In our work, the spectral index $\beta$ around $\nu$ is estimated with
\begin{equation}\label{Eq:Beta_Cal}
\beta=-\frac{\log({F_{1.1\nu}})-\log({F_{\nu/1.1}})}{\log(1.1\nu)-\log(\nu/1.1)}.
\end{equation}
It should be noted that $\beta$ is different from the spectral index $\beta_{0.3-10\rm keV}$,
which is estimated based on the $0.3-10\rm keV$ observations of the X-ray telescope onboard the \emph{Swift} mission.
The value of $\beta_{0.3-10\rm keV}$ may approximate to the value of $\beta$ estimated around $\nu=1.7{\rm keV}/h$,
where $h$ is Planck's constant.
Equation~(\ref{Eq:Beta_Cal}) reveals that the value of $-\beta$ is the spectral slope of intrinsic spectrum
(in the space of $\log \nu'-\log H'_{\nu'}$) around frequency $\nu'=\nu/D$.
Then, we can have
\begin{equation}\label{Eq:for alpha}
\log {H'_{\nu/D_0}} - \log {H'_{\nu/D}} = \int_{\log(\nu /{D_0})}^{\log(\nu /D)} {\beta d[\log(\nu/D)]}.
\end{equation}
With Equations~(\ref{Eq:Observed flux}), (\ref{Eq:D}), (\ref{Eq:Fnu_Evolution}) and $t_c=t_{c,r}$, we can have
\begin{equation}\label{ApA:alpha_beta_relation}
\alpha(t_{\rm obs})=2+\frac{1}{\log (1 + t_{\rm obs}/t_c)}\int_{0}^{{\log (1 + t_{\rm obs}/t_c)}} {\beta(\tau)d[\log (1 + \tau/t_c)]}.
\end{equation}
This is the relation of $\alpha$ and $\beta$.
For $\beta(t_{\rm obs})=\rm constant$, the relation of $\alpha=2+\beta$ can be found from Equation~(\ref{ApA:alpha_beta_relation}).

\section{Discussion about the value of $\tilde{t}_c$}\label{Sec:tc_discussion}
In this section, we discuss the value of $\tilde{t}_c$ under the situation that
$t_0$ is set at a certain time in the steep decay phase.
Based on the analysis in Section~\ref{Sec:Analytical Formula of Flux Evolution},
$\tilde{t}_c=t_{c,r}+t_p^0$ can be found for an EFCS located at $r$.
However, the value of $\tilde{t}_c$ for a jet shell radiating from $r_0$ to $r_e$
depends on the behavior of jet's dynamics and radiation.
In this section, we discuss the situation in Section~\ref{Sec:reality} with Case (I) and $k=0$.

As showed in Section~\ref{Sec:Numerical calculation Procedure},
an expanding jet in our numerical calculations is modelled with a series of jet shell
at radius $r_0,\;r_1=r_0+\beta_{\rm jet}c{\delta t},\;r_2=r_1+\beta_{\rm jet}c{\delta t},\;\cdot\cdot\cdot,r_n=r_{n-1}+\beta_{\rm jet}c{\delta t},\;\cdot\cdot\cdot$
appearing at the time $t=0,\;{\delta t},\;{2\delta t},\;\cdot\cdot\cdot, n{\delta t},\;\cdot\cdot\cdot $
with velocity $\beta_{\rm jet}(r_0),\;\beta_{\rm jet}(r_1),\;\beta_{\rm jet}(r_2),\;\cdot\cdot\cdot,\beta_{\rm jet}(r_n),\;\cdot\cdot\cdot$, respectively.
During the shell's expansion for $\delta t$,
the shell move from $r_{n-1}$ to $r_n$ with the same radiation behavior for emitters.
That is to say, the radiation of our jet can be regarded
as the radiation from a series of EFCSs
appearing with different observer time $t_{{\rm obs},r_n}$ and $t_{c,r_n}$.
Then, the observed flux $F_{\nu}(\tilde{t}_{\rm obs})$ can be described as
\begin{equation}\label{Eq:ApB_F_nu_sum}
F_\nu(\tilde{t}_{\rm obs})=\sum_{r_n}{F_{\nu,r_n}(\tilde{t}_{\rm obs})}
\end{equation}
with $F_{\nu,r_n}(\tilde{t}_{\rm obs})$ being the observed flux at $\tilde{t}_{\rm obs}$ from the EFCS located at $r_n$.
According to the discussion in Section~\ref{Sec:Analytical Formula of Flux Evolution},
one can have
\begin{equation}\label{Eq:ApB_F_nu_rn}
F_{\nu,r_n}(\tilde{t}_{\rm obs})
=\tilde{F}_{\nu,r_n,0}\left(1 + \frac{\tilde{t}_{\rm obs}}{t_{c,r_n}+{t_0}-t_{{\rm obs}, r_n}}\right)^{ - 2-\hat{\beta}'_0},
\end{equation}
where $\tilde{F}_{\nu,r_n,0}$ is the observed flux at $\tilde{t}_{\rm obs}=0$ for the EFCS located at $r_n$,
and ${t_{{\rm{obs}},{r_n}}}$ is the observed time for the first photon from this EFCS.
With Equations~(\ref{Eq:t_obs_r}) and (\ref{Eq:Gamma_Evolution}), one can have
\begin{equation}
{t_{{\rm{obs}},{r_n}}}
\approx {\int_{r_0}^{r_n} \frac{1}{2\Gamma ^2c}dl} =
\left\{ {\begin{array}{*{20}{c}}
{({{t_{c,{r_n}}} - {t_{c,r_0}}} )/(1 - 2s),}&{s \ne 0.5,}\\
{t_{c,r_0}\ln ( {{r_n/r_0}} ),}&{s = 0.5.}
\end{array}} \right.
\end{equation}
For significantly large value of $\Gamma\;(\gtrsim 5)$,
the difference between ${t_{{\rm{obs}},{r_n}}}$ and $\int_{{r_0}}^{{r_n}} {{{(2{\Gamma ^2}c)}^{ - 1}}} dr$
can be neglected. Then, we take ${t_{{\rm{obs}},{r_n}}}=\int_{{r_0}}^{{r_n}} {{{(2{\Gamma ^2}c)}^{ - 1}}} dr$.
In addition, $t_{c,r_n}=t_{c,r_0}$ for $s=0.5$ can be found based on Equation~(\ref{Eq:Gamma_Evolution}).
Thus, one can have following relation:
\begin{equation}\label{Eq:tc_Evolution}
t_{c,r_n}=t_{c,r_0}+(1-2s)t_{{\rm obs},r_n}.
\end{equation}
Accordingly, Equation~(\ref{Eq:ApB_F_nu_sum}) can be reduced to
\begin{equation}\label{Eq:ApB_F_nu}
F_\nu(\tilde{t}_{\rm obs})=\sum_{r_n}{\tilde{F}_{\nu,r_n,0}
\left(1 + \frac{\tilde{t}_{\rm obs}}{t_{c,r_0}-2st_{{\rm obs}, r_n}+{t_0}}\right)^{ - 2-\hat{\beta}'_0}}=\tilde{F}_{\nu ,0}{\left(1 + \frac{\tilde{t}_{\rm obs}}{\tilde{t}_c}\right)^{- 2-\hat{\beta}'_0}},
\end{equation}
which describes the flux evolution in the steep decay phase.
Different value of $s$ may form different value of $\tilde{t}_c$.

For $s=0$, the value of $t_{c,r_0}-2st_{{\rm obs}, r_n}+{t_0}=t_{c,r_0}+{t_0}$ is the same one for different $F_{\nu,r_n}$.
That is to say, for different $r_n$ the dependence of $F_{\nu,r_n}(\tilde{t}_{\rm obs})$ on $\tilde{t}_{\rm obs}$
is the same except the difference in $\tilde{F}_{\nu,r_n,0}$.
Then, one can find $\tilde{t}_c=t_{c,r_0}+{t_0}$,
which is consistent with those found by fitting the flux in the middle panel of Figure~4.
For $s\neq 0$, the value of $t_{c,r_0}-2st_{{\rm obs}, r_n}+{t_0}$ depends on the value of $t_{{\rm obs}, r_n}$.
For $s=-1$, the decay timescale of $F_{\nu,r_n}$ increases with $t_{{\rm obs},r_n}$.
The maximum and minimum decay timescales of $F_{\nu,r_n}$
are $t_{c,r_{\rm e}}+{t_0}-t_p$ and $t_{c,r_0}+{t_0}$, respectively.
Here, $t_{{\rm obs},r_{\rm e}}=t_p$ is found in our numerical calculations.
Based on Equation~(\ref{Eq:ApB_F_nu}),
one can find the relation of $3.32t_{c,r_0} <\tilde t_{c}<8t_{c,r_0}$ if ${t_0}=t_p$ is set,
where $t_{c,r_{\rm e}}=8t_{c,r_0}$ is estimated based on Equation~(\ref{Eq:Gamma_Evolution}) and $s=-1$.
The reported result by fitting the data in the left panel of Figure~4,
i.e., $\tilde t_{c}=6.41t_{c,r_0}$ for ${t_0}=t_p$, is consistent with above analysis.
For $s=1$, the flux decay timescale would decrease with increasing $t_{{\rm obs},r_n}$.
Then, the maximum and minimum decay timescales of $F_{\nu,r_n}$
are $t_{c,r_0}+{t_0}$ and $t_{c,r_{\rm e}}+{t_0}-t_p$, respectively.
According to Equation~(\ref{Eq:ApB_F_nu}),
one should find the relation of $0.5t_{c,r_0}<\tilde t_{c}<1.50t_{c,r_0}$ if ${t_0}=t_p$ is set,
where $t_{c, r_{\rm e}}=0.5t_{c,r_0}$ is estimated based on Equation~(\ref{Eq:Gamma_Evolution}) and $s=1$.
The fitting result from the right panel of Figure~4,
i.e., $\tilde t_c=0.69t_{c,r_0}$ for ${t_0}=t_p$, also confirms our analysis.
Since $\tilde t_{c}=6.41t_{c,r_0}\sim 8t_{c,r_0}$ ($\tilde t_c=0.69t_{c,r_0}\sim 0.5t_{c,r_0}$) is found for $s=-1$ ($s=1$),
the flux of the steep decay phase in our numerical calculations is dominated by the emission from the EFCS located at $\sim r_e$.
Then, we use the decay timescale found in Case (I) with $k=0$, i.e.,
$6.41t_{c,r_0}$, $1.96t_{c,r_0}$, and $0.69t_{c,r_0}$,
to discuss the situations with Cases (II) or (III).



\clearpage
\begin{figure}\label{Fig:Thin_Shell_Case_I}
\plotone{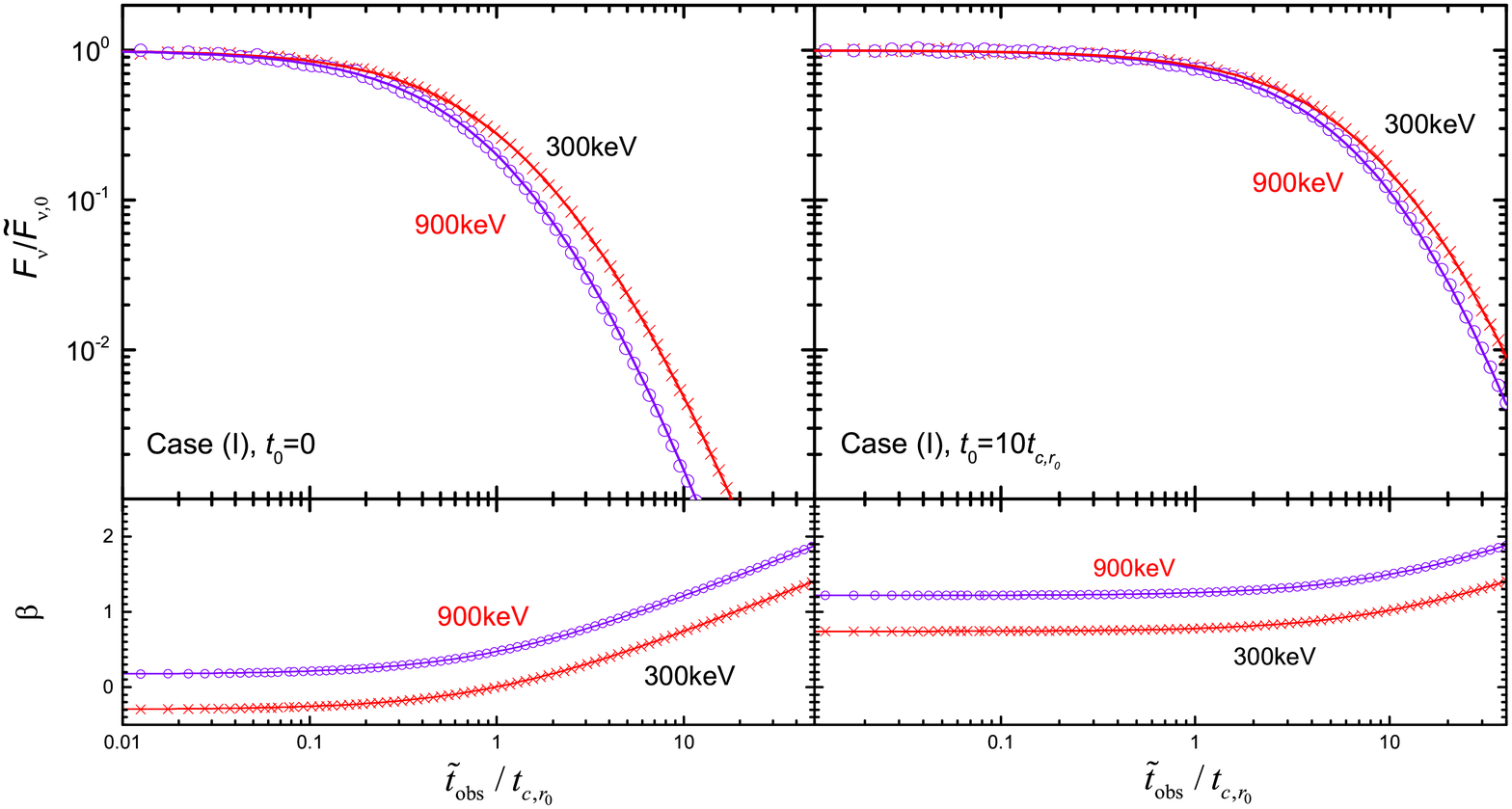}
\caption{
Evolutions of flux (upper panels) and spectral indexes (lower panels)
for an EFCS in Case (I) based on our numerical calculations.
The red ``$\times$'' and  violet ``$\circ$'' represent the data for the
observed photon energy $h\nu=300{\rm keV}$ and $900{\rm keV}$, respectively.
The red (violet) lines describe Equations~(21) and (23)
in the upper and lower panels for $h\nu=300{\rm keV}$ ($900{\rm keV}$), respectively.
Here, $\tilde{t}_c=t_{c,r_0}$ and $11t_{c,r_0}$ are adopted in the left and right panels, respectively.
}
\end{figure}

\clearpage
\begin{figure}\label{Fig:Thin_Shell_Case(II)&(III)}
\plotone{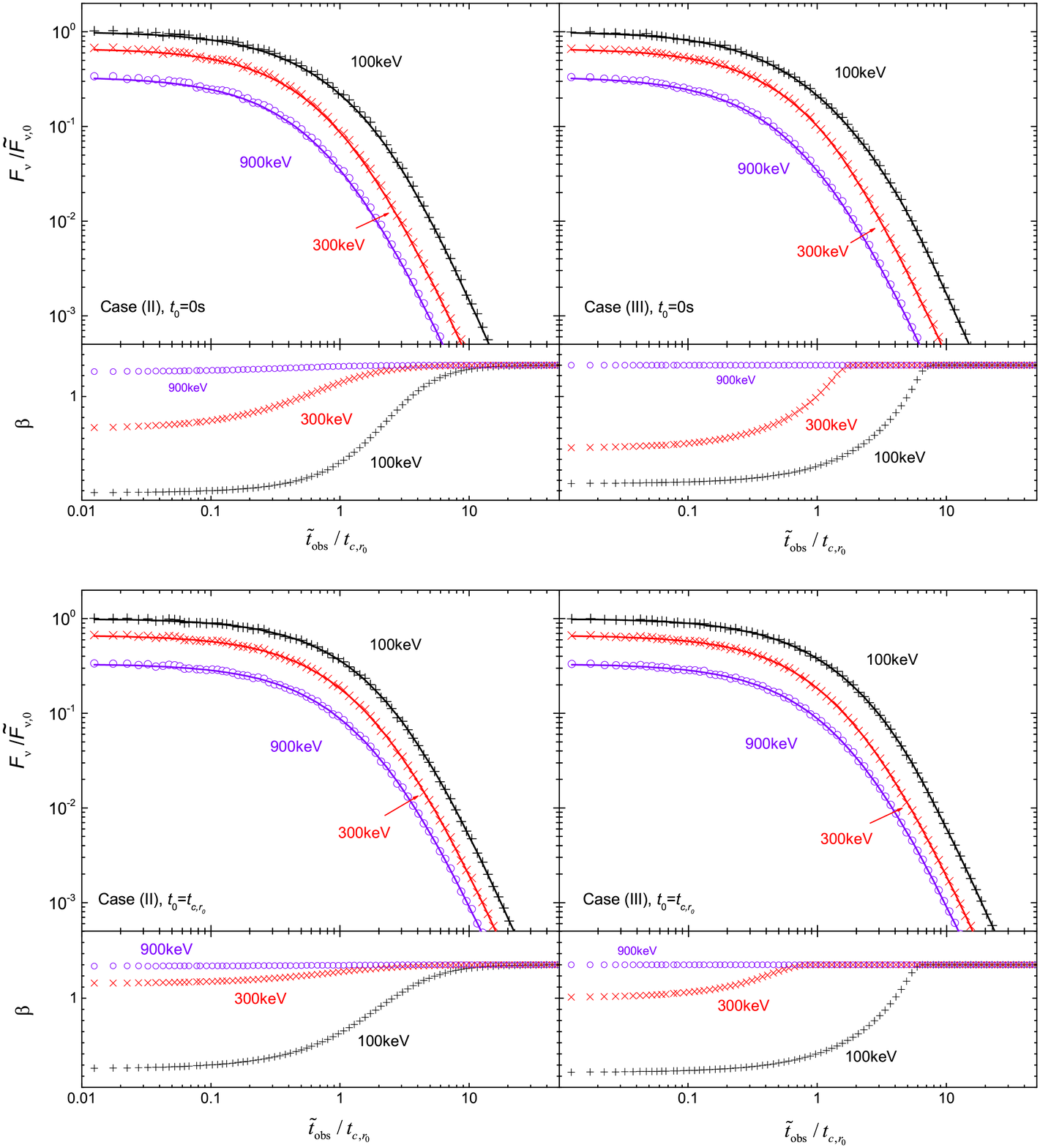}
\caption{
Evolutions of flux (upper panels) and spectral indexes (lower panels)
for an EFCS in Case (II) (left panels) or (III) (right panels).
The black ``+'', red ``$\times$'', and violet ``$\circ$'' represent the data for the
observed photon energy $h\nu=100{\rm keV}$, $300{\rm keV}$, and $900{\rm keV}$, respectively.
The black, red, and violet lines represent the flux following Equation~(21)
for $h\nu=100{\rm keV}$, $300{\rm keV}$, and $900{\rm keV}$ respectively,
where $\tilde{t}_c=t_{c,r_0}$ and $\tilde{t}_c=2t_{c,r_0}$ are adopted in upper panels and lower panels, respectively.
The light curves for $h\nu=300{\rm keV}$ ($900{\rm keV}$)
are shifted by dividing $1.5$ ($3$) in the plot for clarity.
}
\end{figure}

\clearpage
\begin{figure}\label{Fig_Thin_Shell_Case_(IV)}
\plotone{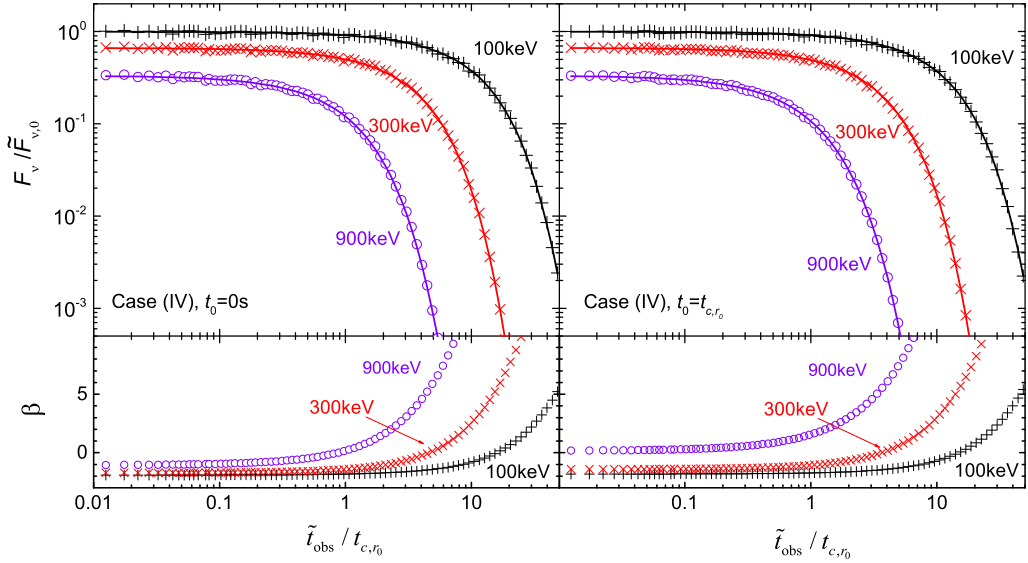}
\caption{
Evolutions of flux (upper panels) and spectral indexes (low panels) for an EFCS with Case (IV).
The meanings of symbols and solid lines are the same as those in Figure~2,
where the value of $t_0=0\rm s$, $\tilde t_c=t_{c,r_0}$ and $t_0=t_{c,r_0}$, $\tilde t_c=2t_{c,r_0}$ are adopted in the left and right panels, respectively.
}
\end{figure}

\clearpage
\begin{figure}\label{Fig:Real_Case_I_and_k_0}
\plotone{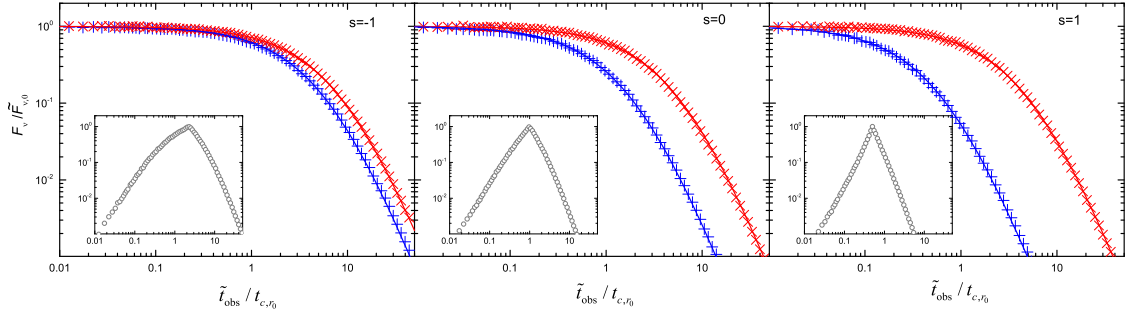}
\caption{
Flux evolution for an expanding thin jet shell with Case (I) and $k=0$.
The gray ``$\circ$'', blue ``+'', and red ``$\times$'' represent the data
by setting $t_0=0\rm s$, $t_p$ and $5t_{c,r_0}$, respectively.
The blue and red lines are the best fitting results about flux evolution with Equation~(21) and $\alpha=3.3$.
}
\end{figure}

\clearpage
\begin{figure*}\label{Fig:Real_Case_II}
\plotone{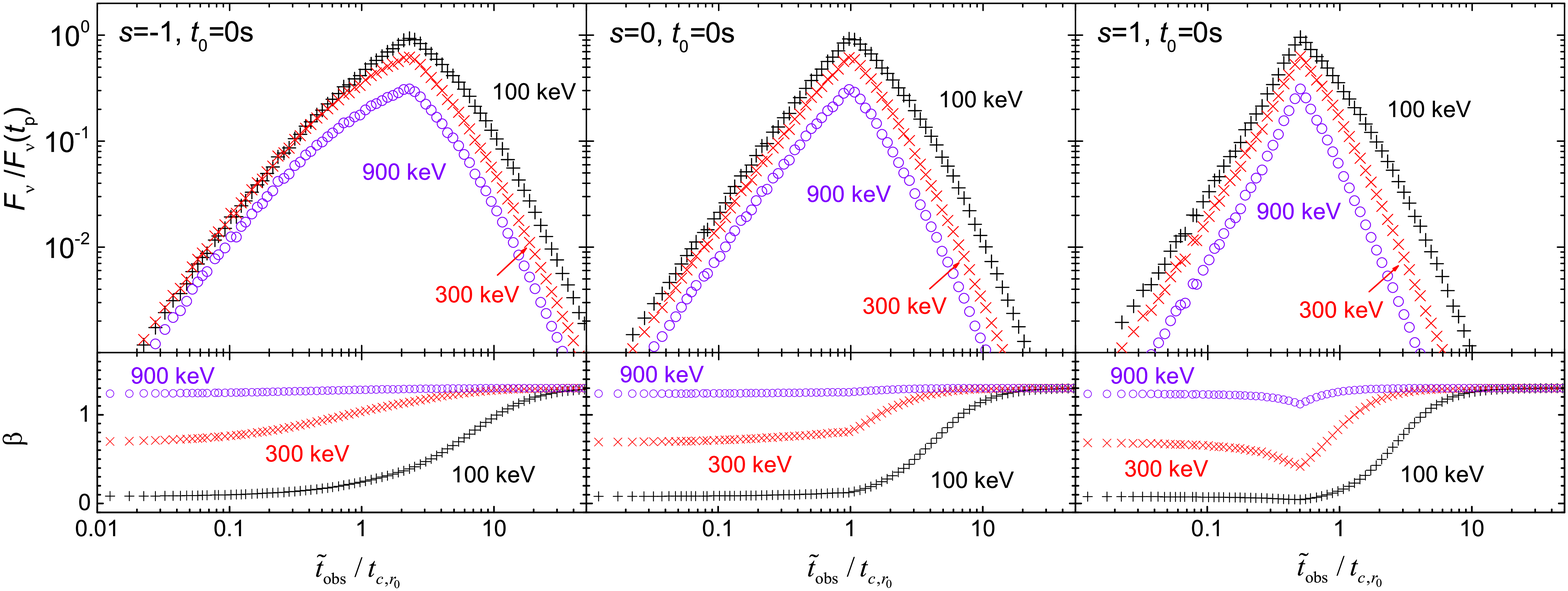}
\plotone{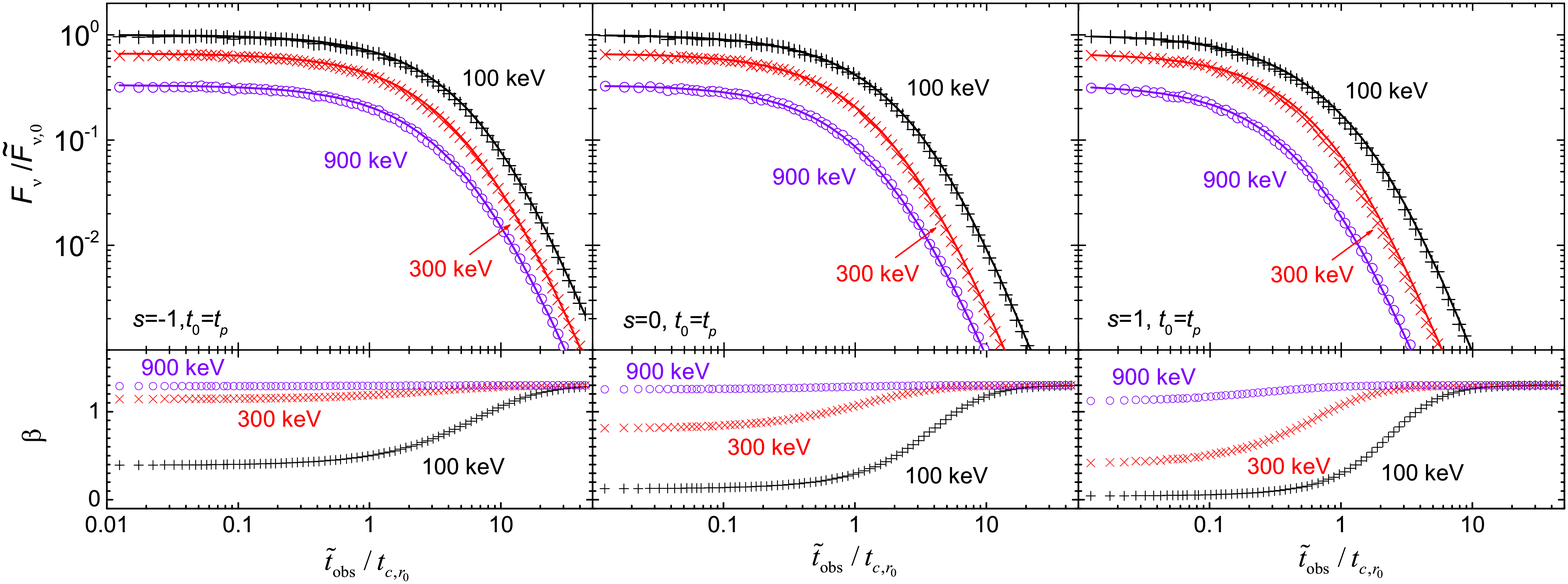}
\plotone{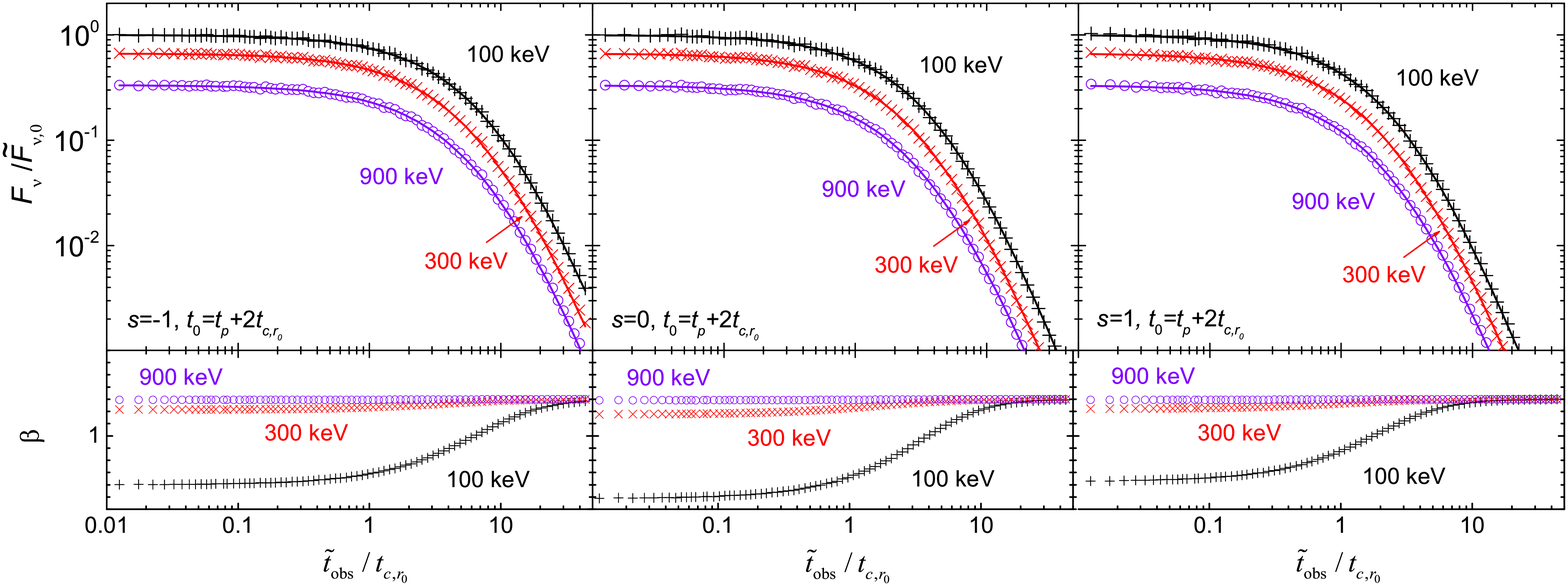}
\caption{
Flux and spectral evolution in real situation for Case (II).
The meaning of symbols and solid lines are the same as those in Figure~2,
where the value of $\tilde t_{c}=6.41t_{c,r_0}$ ($8.41t_{c,r_0}$), $1.96t_{c,r_0}$ ($3.96t_{c,r_0}$), and $0.69t_{c,r_0}$ ($2.69t_{c,r_0}$) are adopted in Equation (21) for $s=-1$, $0$, and $1$ with $t_0=t_p$ ($t_p+2t_{c,r_0}$), respectively.
}
\end{figure*}

\clearpage
\begin{figure*}\label{Fig:Real_Case_III}
\plotone{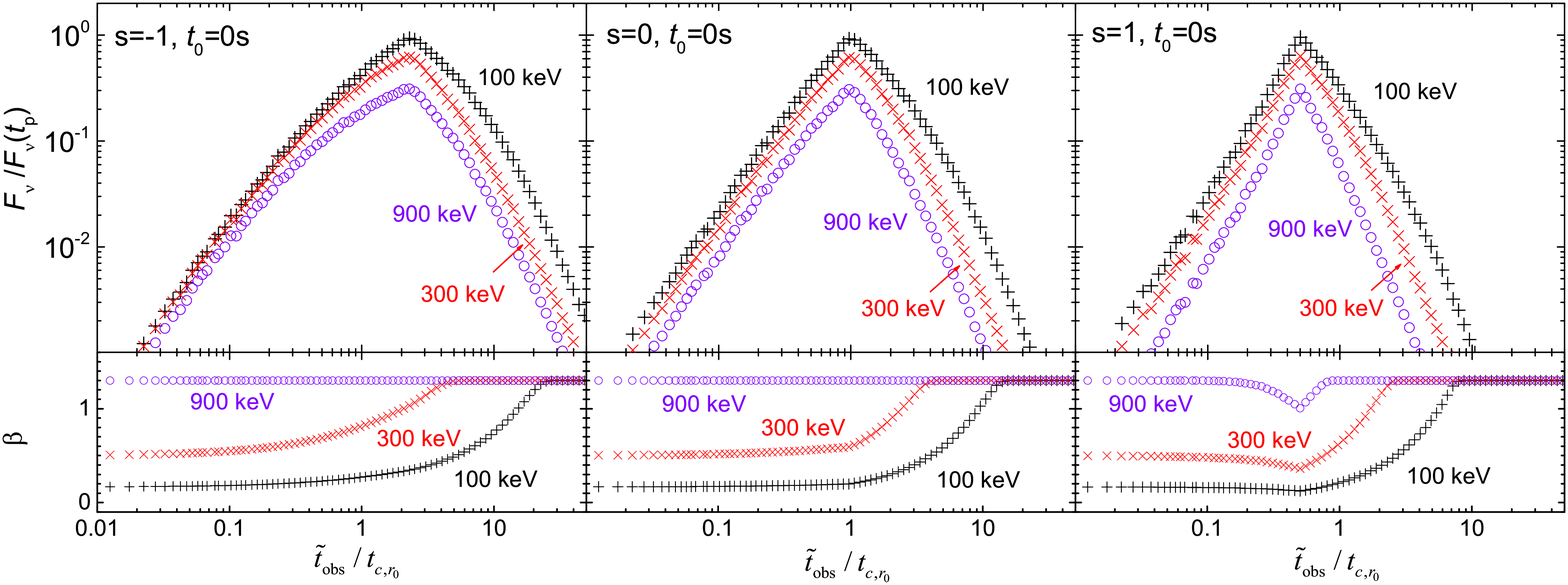}
\plotone{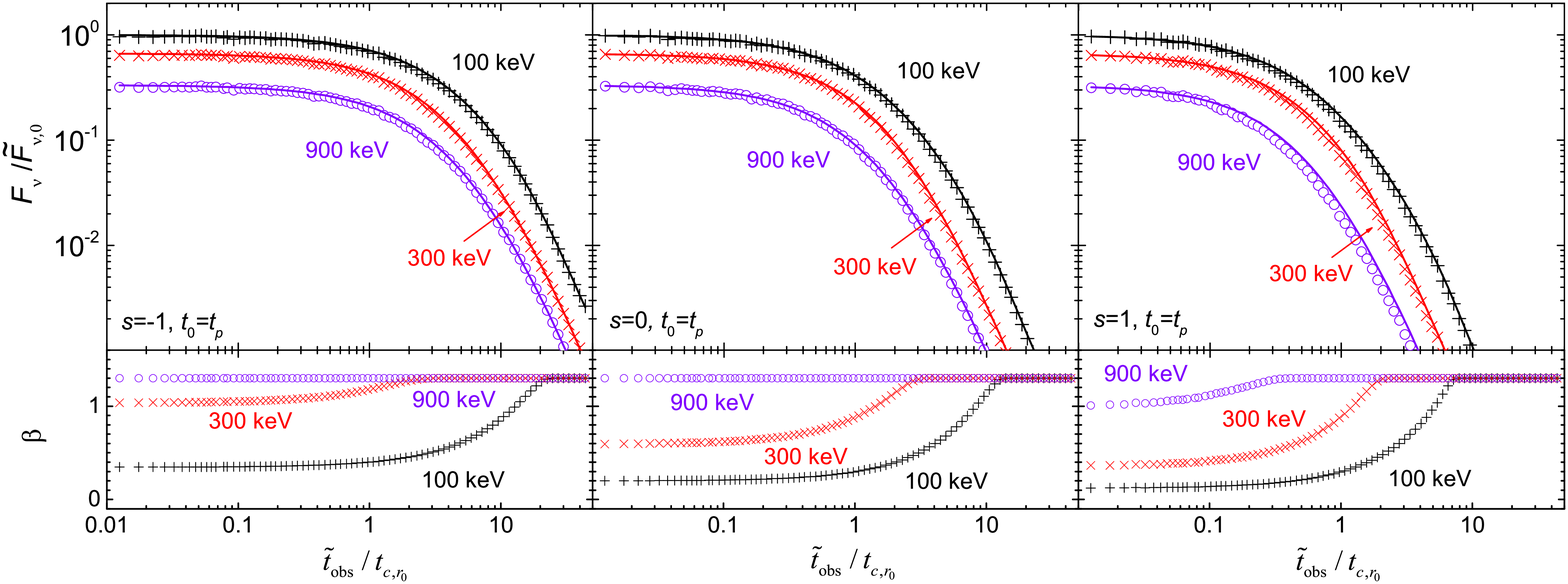}
\plotone{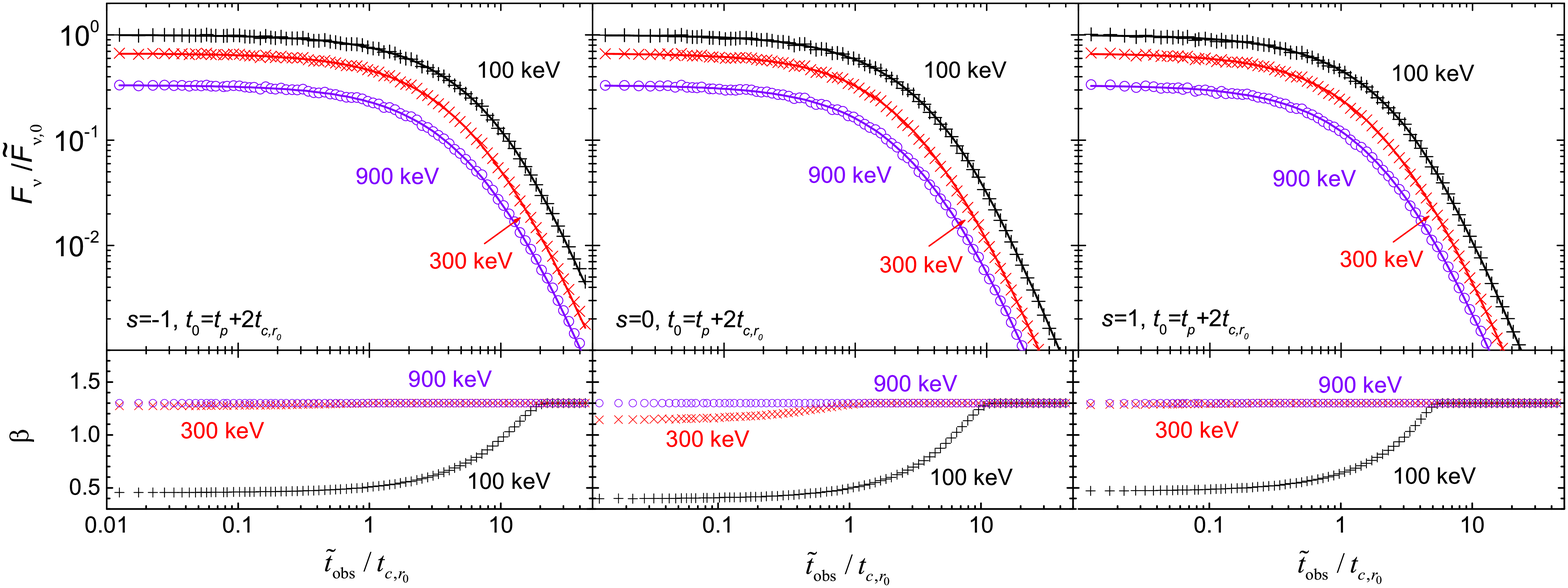}
\caption{
Flux and spectral evolution in real situation for Case (III).
The meanings of symbols and solid lines are the same as those in Figure~5.
}
\end{figure*}

\clearpage
\begin{figure*}\label{Fig:Real_Case_II_diffalphabeta}
\plotone{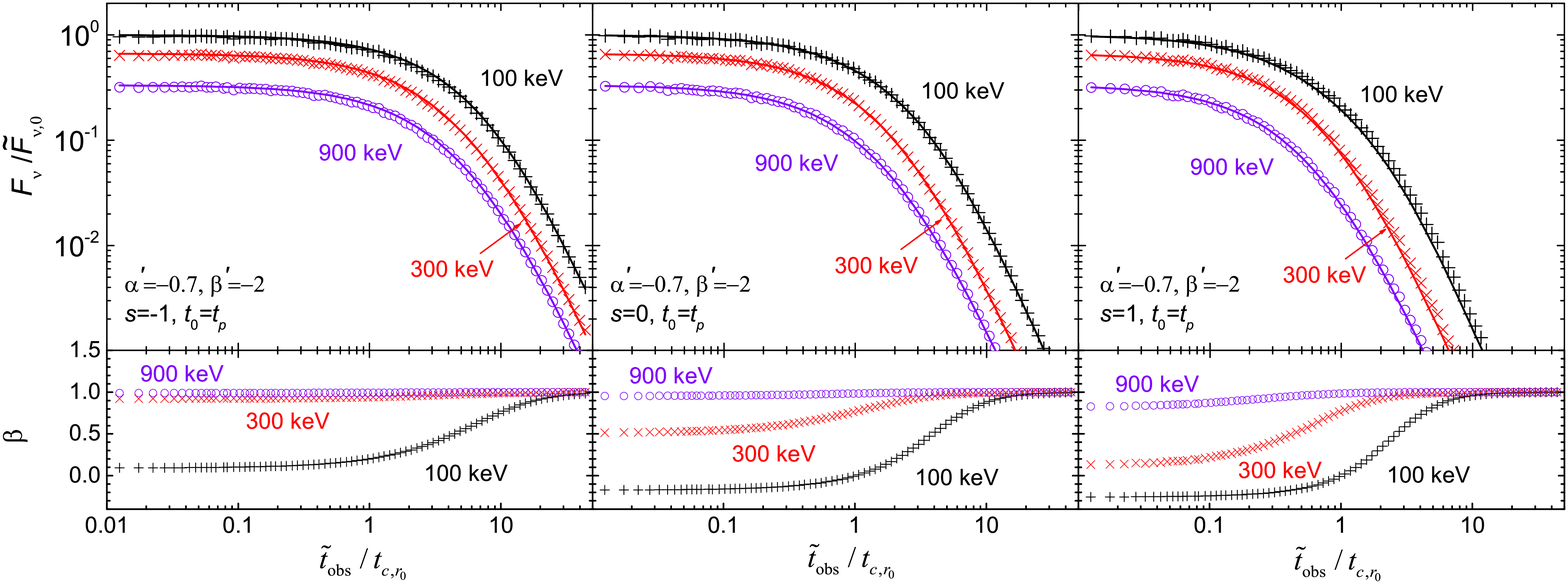}
\plotone{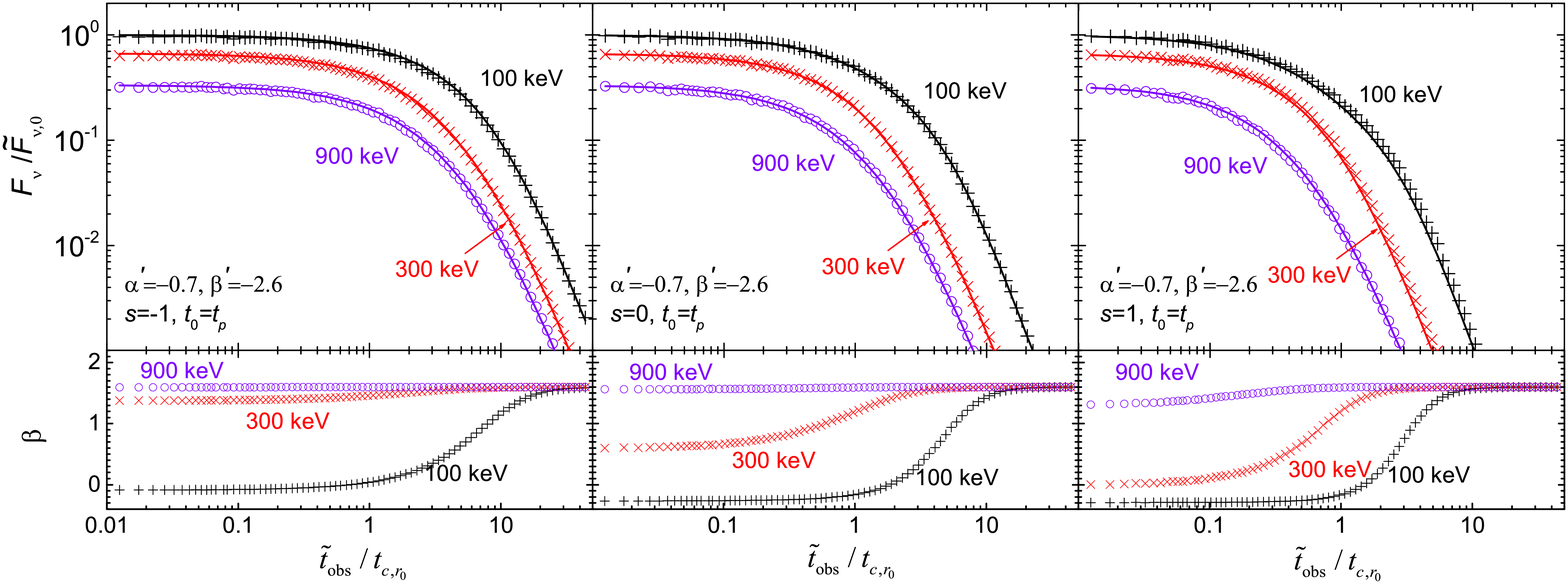}
\plotone{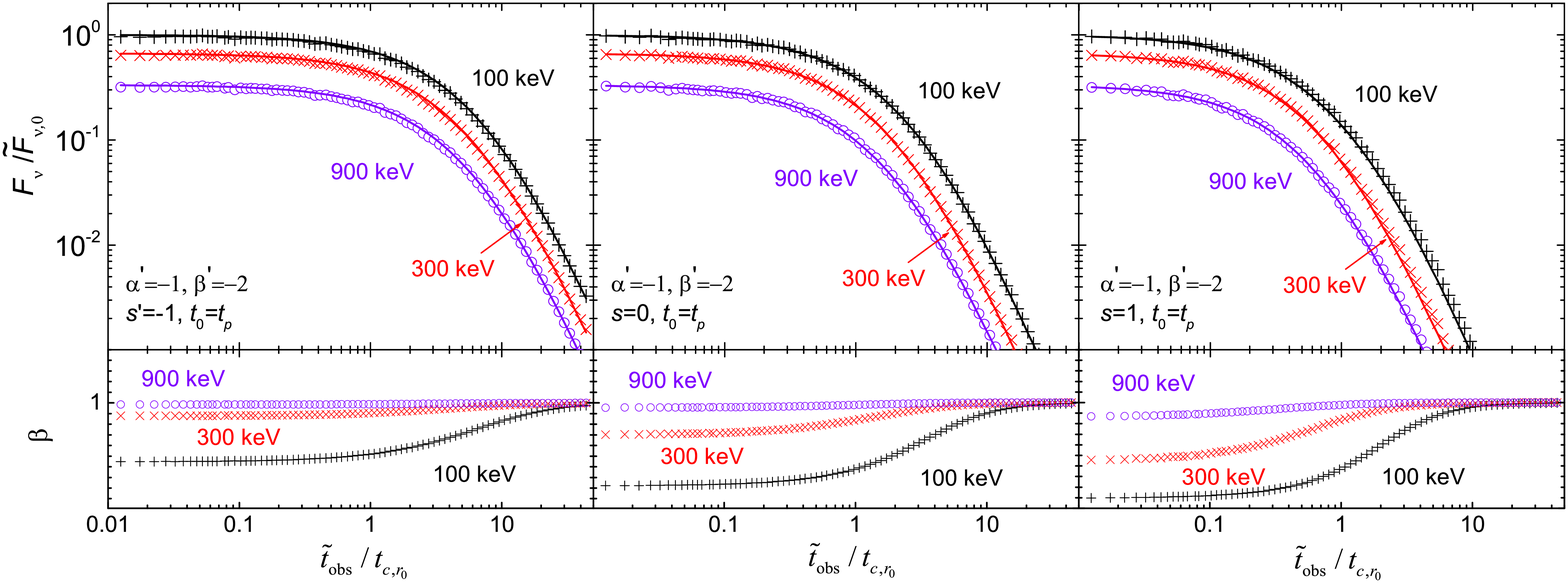}
\caption{
Flux and spectral evolution in real situation for Case (II)
with different value of $\alpha'$ and $\beta'$.
The meanings of symbols and solid lines are the same as those in Figure~5.
}
\end{figure*}
\addtocounter{figure}{-1}
\begin{figure*}
\plotone{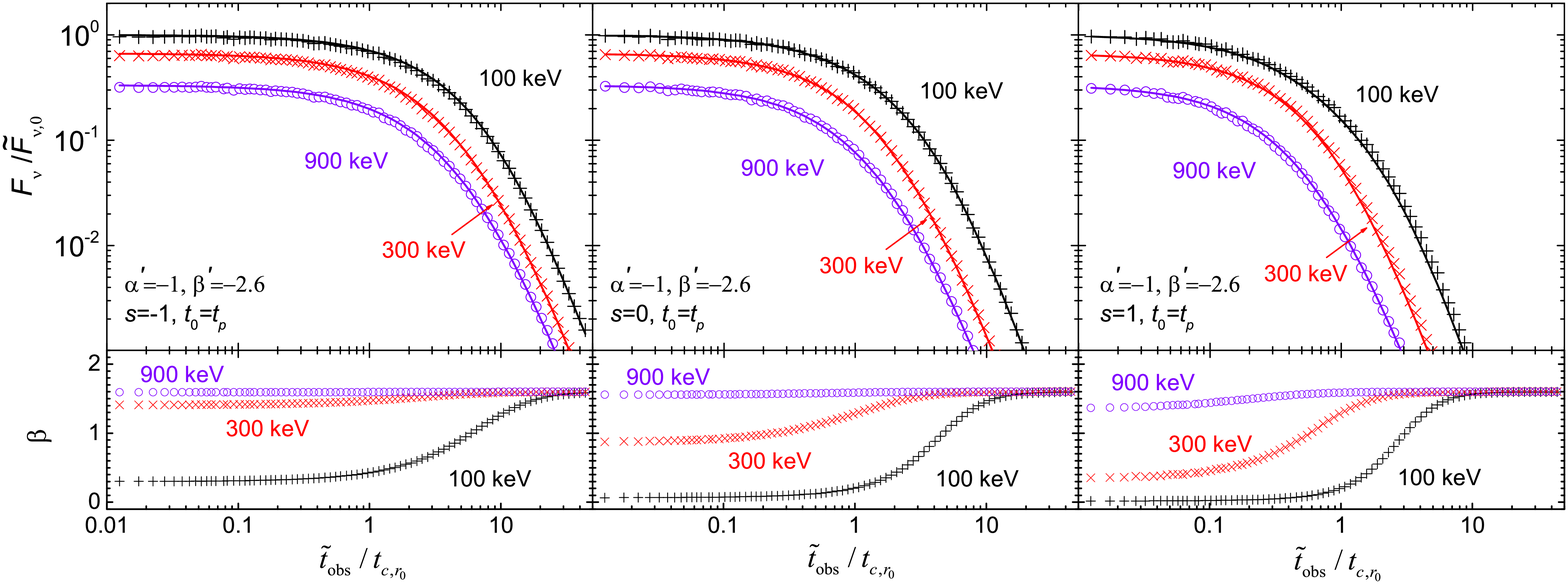}
\caption{(\emph{Continued})}
\end{figure*}

\clearpage
\begin{figure*}\label{Fig:Real_Case_III_diffalphabeta}
\plotone{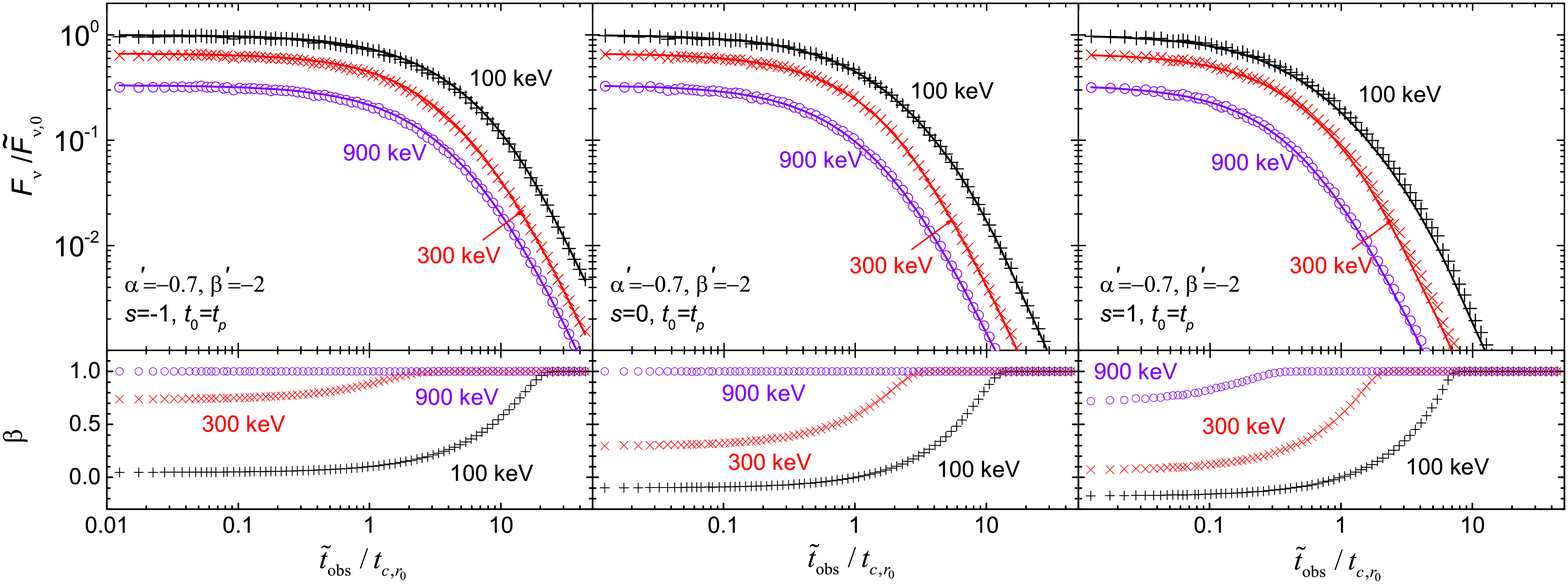}
\plotone{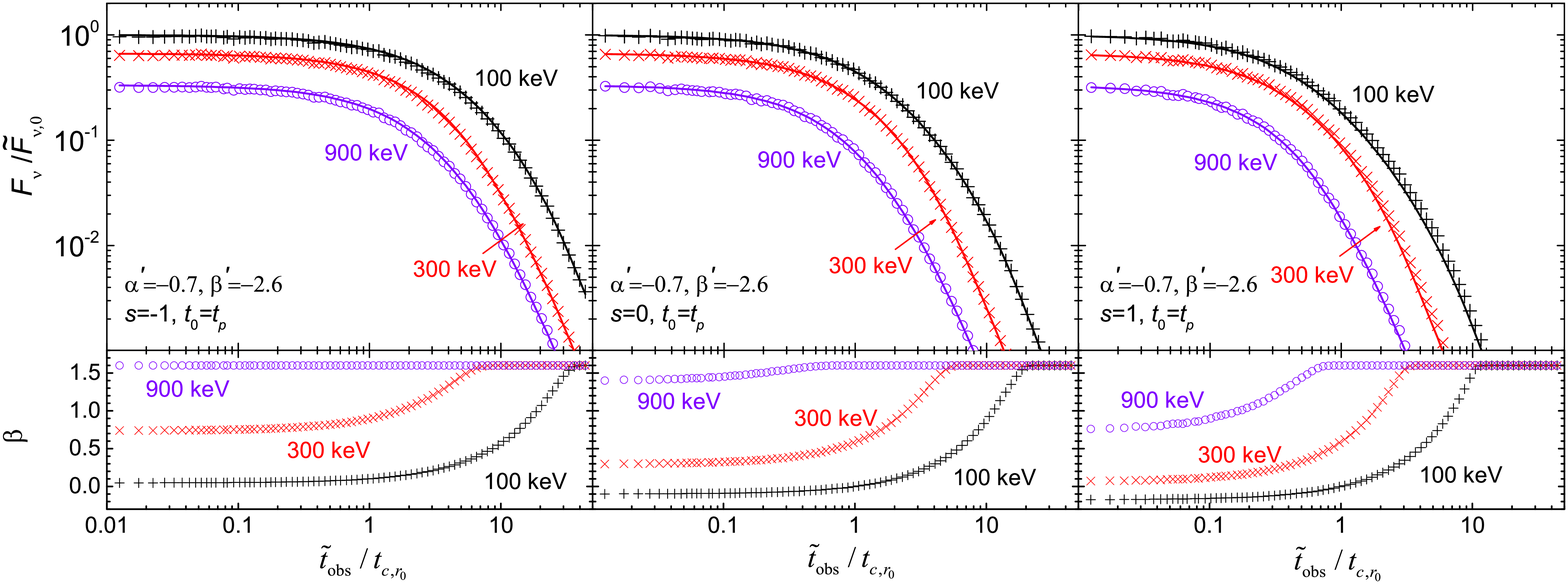}
\plotone{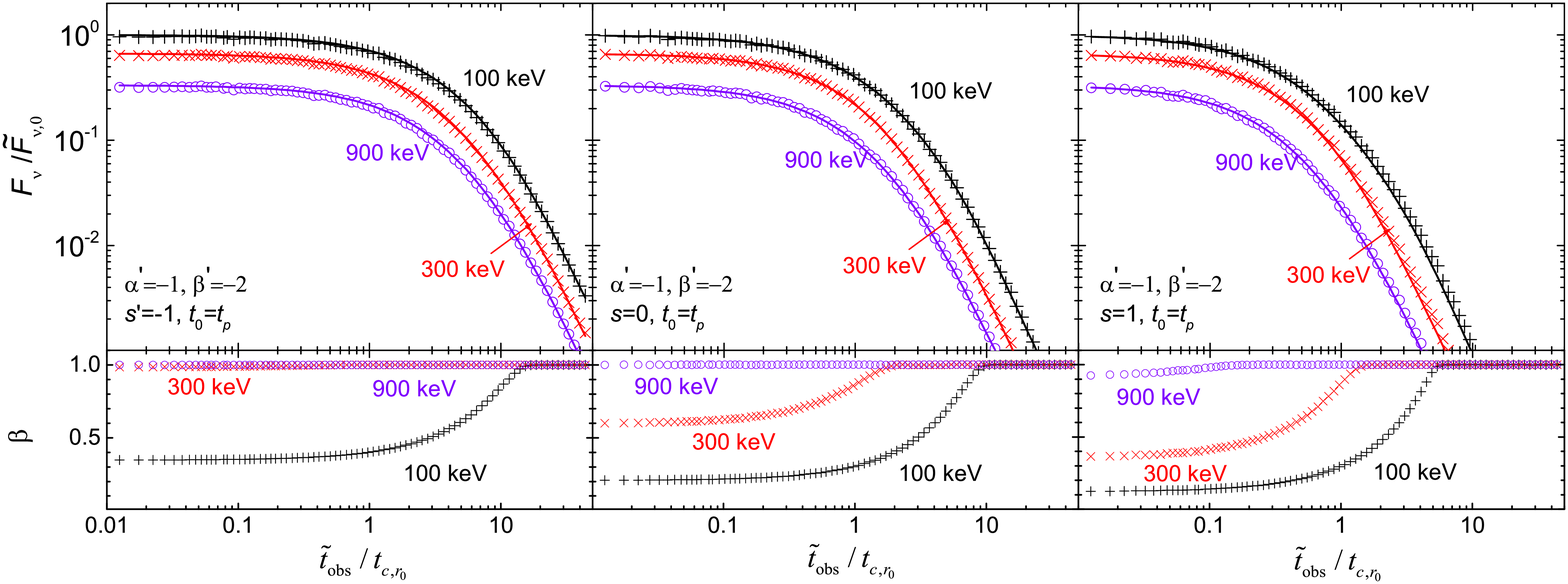}
\caption{
Flux and spectral evolution in real situation for Case (III)
with different value of $\alpha'$ and $\beta'$.
The meanings of symbols and solid lines are the same as those in Figure~5.
}
\end{figure*}
\addtocounter{figure}{-1}
\begin{figure*}
\plotone{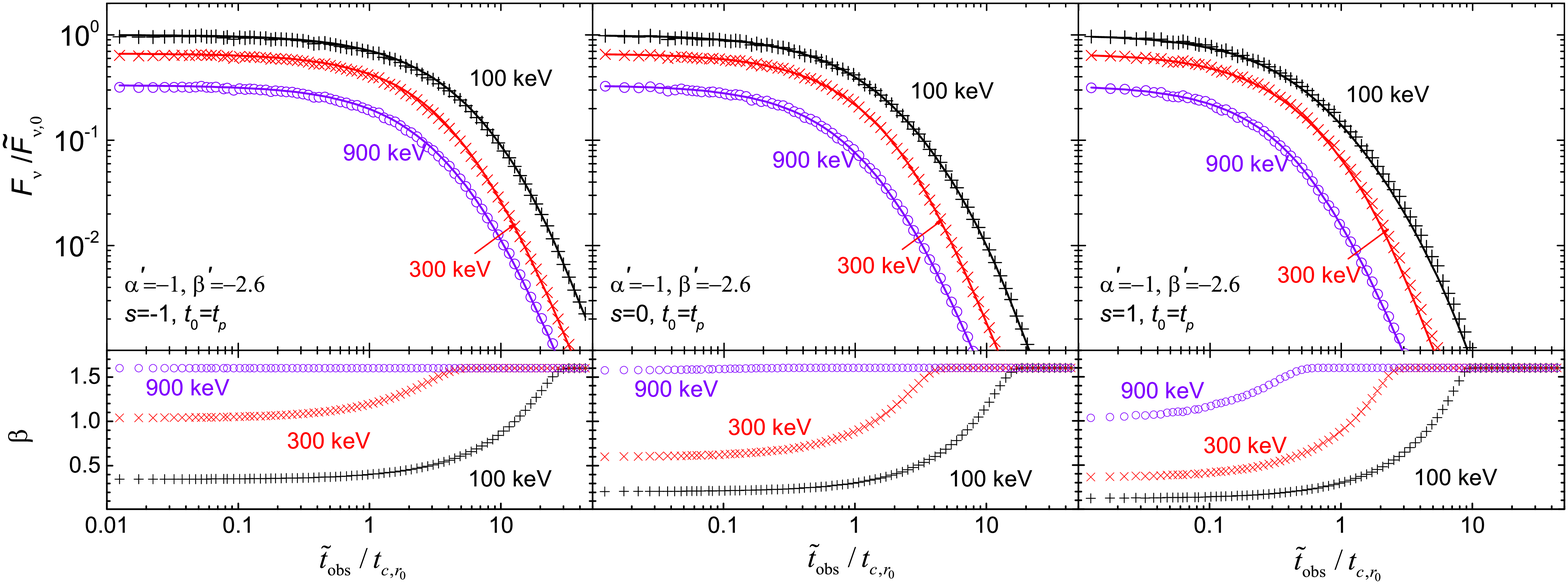}
\caption{(\emph{Continued})}
\end{figure*}
\end{document}